\documentclass[10pt]{iopart}

\eqnobysec 

\renewcommand{\vec}[1]{\mbox{\boldmath $#1$}}  
\newcommand\beq{\begin{equation}}
\newcommand\eeq{\end{equation}}
\newcommand\bea{\begin{eqnarray}}
\newcommand\eea{\end{eqnarray}}

\newcommand\pa{\partial}
\newcommand\prandtl{P}
\newcommand{\Jsn}{\mbox{ sn}}
\newcommand{\Jcn}{\mbox{ cn}}
\newcommand{\Jdn}{\mbox{ dn}}
\newcommand{\sech}{\mbox{ sech}}

\def\d#1#2{{{\rm d} #1 \over {\rm d} #2}}

\def\prt#1#2{{\pa #1 \over \pa #2}}

\newcommand\half{{\textstyle \frac12 }}
\newcommand\tsfrac[2]{{\textstyle \frac{#1}{#2} }}

\newcommand\ep{\epsilon}
\renewcommand\i{{\rm i}}

\input epsf

\begin{document}

\title{Pattern formation with a conservation law}

\author{P. C. Matthews and S. M. Cox}

\address{School of Mathematical Sciences,
University of Nottingham, \\
University Park,
Nottingham NG7 2RD, UK
}

\begin{abstract}
Pattern formation in systems with a conserved quantity is
considered by studying the appropriate amplitude equations.
The conservation law leads to a large-scale neutral mode
that must be included in the asymptotic analysis for pattern formation
near onset. Near a stationary bifurcation,
the usual Ginzburg--Landau equation for the amplitude of
the pattern is then coupled to an equation for the large-scale mode.
These amplitude equations show that for certain parameters all roll-type
solutions are unstable.  This new  instability differs from the Eckhaus
instability in that it is amplitude-driven and is supercritical.
Beyond the stability boundary, there exist stable
stationary solutions in the form of strongly modulated patterns.
The envelope of these modulations is calculated in terms of Jacobi
elliptic functions and, away from the onset of modulation, is
closely approximated by a sech profile. Numerical simulations
indicate that as the modulation becomes more pronounced, the envelope
broadens.
A number of applications are considered, including convection
with fixed-flux boundaries and convection in a magnetic field,
resulting in new instabilities for these systems. 
\end{abstract}


\section{Introduction}

The mathematical theory of pattern formation has a wide range of
applications. In the field of fluid mechanics, the most widely studied
example is Rayleigh--B\'enard convection and its  variants,
but there are many other examples such as Taylor--Couette flow and
solidification of alloys. Pattern-forming systems can also be found
in many chemical and biological systems (see the review by Cross and
Hohenberg 1993).

In this paper we consider pattern formation in one space dimension in
systems sufficiently large to allow a large-scale modulation of the
pattern. It is assumed that the system has some reflection symmetry,
that the onset of instability is stationary, and that the onset of
pattern formation occurs with a non-zero wavenumber.
In this case, the equation governing the modulation
of the small-amplitude pattern is in general the real
Ginzburg--Landau equation.
This well known result has recently been given rigorously by Melbourne (1999).
The behaviour of the real Ginzburg--Landau equation is straightforward:
a regular periodic pattern is stable provided its wavenumber lies
within the Eckhaus band (Eckhaus 1965); patterns outside this band
are unstable, and the instability is subcritical 
(Fauve 1987, Tuckerman and Barkley 1990).
Physically,
patterns in which the wavenumber is either too large or too small
adjust to a different wavenumber by adding or losing waves.

However, there is a large class of problems in which the onset of
pattern formation at finite wavenumber is not governed by the
Ginzburg--Landau equation.
Such problems arise when the system possesses a conservation law
(Cross and Hohenberg 1993, p.~883). To be specific, we suppose
that this conservation law may be written in the form
\[
\prt{w}{t}+\prt{}{x}f(w)=0,
\]
with appropriate boundary conditions,
from which it follows that $w$ is conserved, that is,
\[
\d{}{t}\int w\,{\rm d}x = 0,
\]
where the integral is taken over the entire spatial domain. The
conservation law leads to a neutral large-scale mode, which must
be included in the dynamics if a correct description of the
behaviour near onset is to be obtained.

The resulting behaviour depends on how the reflection symmetry acts
on the large-scale mode.  The case in which the large-scale mode
changes sign upon reflection arises when the system has Galilean
invariance (Coullet and Fauve 1985). This case was studied by Matthews
and Cox (1999), who showed that all steady patterns are unstable, with the
instability growing more rapidly than the rate of formation of the
pattern. As a consequence, a highly disordered state is found, even near
onset.

In this paper we study the case where the large-scale mode does not
change sign under reflection. Physically, this corresponds to the mode
representing a density rather than a flow.
Similar problems have been considered by Riecke (1992a, 1992b, 1996)
and Herrero and Riecke (1995) for convection in binary mixtures,
but in this case the bifurcation is oscillatory and asymptotically
self-consistent amplitude equations were not obtained.

Experiments demonstrating the effects of coupling between
short-scale pattern modes and large-scale mean flows have been
carried out in a variety of flows. In surface tension-driven
convection, VanHook {\it et al.}\ (1995, 1997) have shown how the
presence of a large-scale surface-deformational mode can lead to
dry spots and high spots. In a two-layer viscous shear flow
(Barthelet and Charru 1998, Charru and Barthelet 1999),
finite-amplitude interfacial waves become unstable through
coupling to a large-scale mode which may be traced back to
conservation of mass of each fluid. In liquid crystals (Hidaka
{\it et al.}\ 1997),
complicated spatio-temporal convection patterns
are traced to the presence of a large-scale mode arising in the
degeneracy of the system.

The paper is organised as follows.
In \S~\ref{sec:pateqs} we derive the governing amplitude equations
by symmetry and scaling arguments and then, in \S~\ref{sec:anal},
analyse their properties. We consider the stability of small-amplitude
solutions, and the behaviour of fully nonlinear solutions of the
amplitude equations, in particular localised states.
In \S~\ref{sec:numsims} we illustrate how the insights we have gained
apply to a model partial differential equation (PDE),
through numerical simulation.
We also demonstrate how the behaviour of the full PDE diverges from
that predicted by the leading-order amplitude equations, as the parameters are
varied, and analyse this divergence in \S~\ref{sec:hot}.
We then describe, in \S~\ref{sec:examples}, three physical
examples in which our amplitude equations arise.
In \S~\ref{sec:disc} we summarise our results.

\section{Derivation of mean-mode and pattern-mode equations}
\label{sec:pateqs}

In this section we derive the governing amplitude equations for
pattern formation with a conservation law, using two different methods.
In \S~\ref{sec:symderive} the equations are deduced from symmetry
considerations, while in \S~\ref{sec:toypde} a model PDE is introduced
and the amplitude equations are derived
by asymptotic expansions near the onset of pattern formation.

\subsection{Derivation of equations from symmetry}
\label{sec:symderive}

Consider a PDE or system of PDEs with a conserved quantity and a basic
zero state.
If a quantity $w(x,t)$ is conserved, it must be possible to write
$\pa w / \pa t$ as the divergence of a flux.
It follows that when the expression for $\pa w / \pa t$  is linearised, there
can be no terms linear in $w$ except those involving $x$-derivatives.
If we assume, furthermore, that the system possesses the reflection symmetry
\beq
x \to -x, \qquad w \to w
\eeq
then only even $x$-derivatives occur.
Thus $w$ is {\em density-like} (a {\em velocity-like}
quantity would satisfy $x \to -x$, $w \to -w$ instead; see Matthews and Cox
(1999), Tribelsky and Velarde (1996)).
On large spatial scales, the
leading linear term is the second derivative $w_{xx}$, so $w(x,t)$ obeys the
diffusion equation and the  growth rate $\lambda$ of modes with
wavenumber $k$ is proportional to $k^2$.
This indicates that large-scale modes are almost neutral, evolving on
a long timescale, and must be included in the analysis of pattern formation.

Suppose that the system supports stationary patterns with a wavenumber $k$,
which we take to be the critical wavenumber according to linear stability
theory; we may set $k=1$ with no loss of generality.
Near the onset of pattern formation the appropriate
ansatz is then
\beq
w(x,t) \sim \tilde{A}(X,T) \e^{\i x } +  \tilde{A}^*(X,T) \e^{-\i x}
+ \tilde{B}(X,T),
\eeq
where $X$ and $T$ are rescaled forms of $x$ and $t$, and $\tilde A$ and
$\tilde B$ are small amplitudes.
Note that the amplitude $\tilde{A}$ is complex, with $A^*$ denoting the
complex conjugate of $A$, but $\tilde{B}$ is real.

The amplitude equations for $\tilde{A}$ and $\tilde{B}$ can be written down
by imposing the symmetries of translation in $x$
\beq
\tilde A \to \tilde A e^{i \chi}, \qquad \tilde B \to \tilde B
\label{eq:shiftsym}
\eeq
and reflection
\beq
X \to -X, \qquad \tilde A \to  \tilde{A}^*, \qquad \tilde B \to \tilde B .
\label{eq:refsym}
\eeq
In the absence of the large-scale mode $\tilde B$, $\tilde A$ obeys the real
Ginzburg--Landau equation. Coupling terms in this equation
must include $\tilde A$ because of (\ref{eq:shiftsym}), so the leading
coupling term in the $\tilde A$ equation that is consistent with the
symmetries is $\tilde A \tilde B$.
The coefficient of this term is forced to be real by (\ref{eq:refsym}).
In the absence of coupling terms, the equation for $\tilde B$ is the diffusion
equation, and (\ref{eq:shiftsym}) forces coupling terms to
involve equal numbers of $\tilde A$ and $\tilde{A}^*$ terms.
Since $\tilde B$ is conserved, all terms in the $\tilde B$ equation
involve derivatives.  The leading coupling term in the $\tilde B$ equation
that is consistent with (\ref{eq:refsym}) is
therefore $(|\tilde A|^2)_{XX}$.

These considerations give rise to amplitude equations of consistent
asymptotic order if we introduce a small parameter $\epsilon$ such that
$\tilde A=\epsilon A$,
$\tilde B=\epsilon^2 B$, $X=\epsilon x$, $T=\epsilon^2 t$.
Furthermore, we take the system to be
a distance of $O(\epsilon^2)$ from the marginal
surface in parameter space. With these scalings,
the amplitude equations truncated to leading order are then
\bea
A_T & = & A + A_{XX} - A|A|^2 - AB  ,             \label{eq:AT}\\
B_T & = & \sigma B_{XX} + \mu (|A|^2)_{XX},       \label{eq:BT}
\eea
where we have assumed that, in the absence of $B$, the bifurcation of the
pattern mode is supercritical.
All coefficients in (\ref{eq:AT}) have been set to unity
by rescaling $T$, $X$, $A$ and $B$.
There are two coefficients, $\sigma$ and $\mu$, in (\ref{eq:BT})
that cannot be removed by rescaling.
We impose the condition $\sigma > 0$ to ensure that the long-wavelength modes
are weakly damped in the absence of $A$, but $\mu$ may have either sign.
Equations analogous to (\ref{eq:AT})--(\ref{eq:BT}) have previously
been derived in the context of the {\em secondary} stability of cellular
patterns (Coullet and Iooss 1990).

\subsection{Derivation of equations from a model PDE}
\label{sec:toypde}

Consider the PDE
\beq
\frac{\partial w}{\partial t}  =  -\frac{\partial^2}{\partial x^2} \left[
r  w - \left(1+ \frac{\partial^2}{\partial x^2} \right)^2 w
           -  s w^2  - w^3  \right],\label{eq:toypde2}
\eeq
where the terms inside the brackets are just those of the
much-studied Swift--Hohenberg equation (Swift and Hohenberg 1977)
with a symmetry-breaking quadratic term (Sakaguchi and Brand 1996,
Matthews 1998).
This equation has the required properties of reflection symmetry and a
conservation law for $w(x,t)$.

\begin{figure}
\epsfxsize0.7\hsize\epsffile{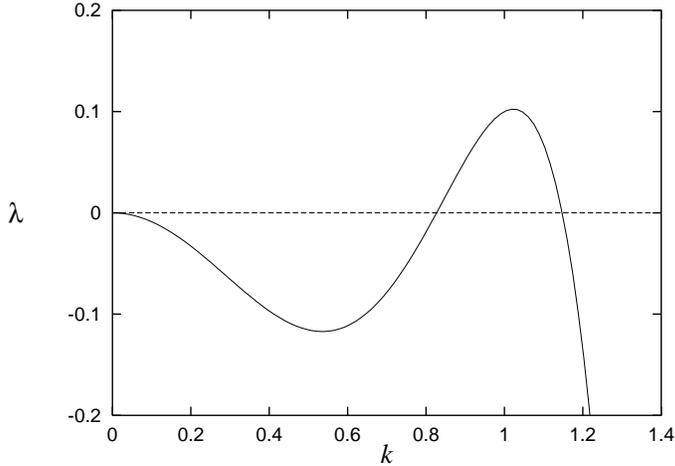}
\caption{The growth rate $\lambda$ of the linear modes of (\ref{eq:toypde2})
as a function of wavenumber $k$, with $r=0.1$.\label{fig:lambda}}
\end{figure}

The growth rate $\lambda$ of modes of wavenumber $k$
in (\ref{eq:toypde2}) is given by
\beq
\lambda = k^2(r  - (1-k^2)^2) ,  \label{eq:lambda}
\eeq
so for $0 < r \ll 1$ a narrow band of wavenumbers near $k=k_c=1$ is
unstable (figure~\ref{fig:lambda}).

We now follow the standard asymptotic methods to describe the
behaviour of (\ref{eq:toypde2}) for small $r$, by writing
\beq
r = \ep^2 r_2, \qquad T = \ep^2 t, \qquad X = \ep x .
\eeq
The appropriate asymptotic expansion for $w(x,t)$ is then
\bea
w(x,t) &=& \ep A(X,T) \e^{\i x} + \ep A^*(X,T) \e^{-\i x} +
\ep^2 B(X,T)\nonumber\\
     & & + \ep^2 C(X,T) \e^{2 \i x} + \ep^2  C^*(X,T) \e^{-2 \i x} + O(\ep^3),
\label{eq:wexp}
\eea
where the large-scale mode $B$ has been introduced at order $\ep^2$.

The expansion (\ref{eq:wexp}) is substituted into (\ref{eq:toypde2})
and the system is then solved at successive orders of $\ep$.
At $O(\ep)$, (\ref{eq:toypde2}) is automatically satisfied.
At $O(\ep^2)$, terms involving $\e^{2 \i x}$ appear, and from these we
deduce that $ C = - sA^2 /9 $.
At $O(\ep^3)$, the amplitude equation for $A$ is deduced from the
terms in $\e^{\i x}$. The equation for $B$ follows from the terms
independent of $x$ at $O(\ep^4)$.
These equations are
\bea
A_T & = & r_2 A + 4 A_{XX} - (3-2s^2/9)|A|^2 A - 2 s AB  , \label{eq:AToy2}\\
B_T & = & B_{XX} + 2 s (|A|^2)_{XX}.       \label{eq:BToy2}
\eea
The sign of $s$ is thus immaterial, and the bifurcation is
supercritical if $s^2 < 27/2$. If this condition is satisfied
then (\ref{eq:AToy2}) and (\ref{eq:BToy2}) can be rescaled to give
(\ref{eq:AT}--\ref{eq:BT}).

Although the system (\ref{eq:AT})--(\ref{eq:BT}) is generic for
pattern-forming systems with a conservation law,
in fact a conservation law alone is not sufficient to yield these equations.
Only the first  of the two nonlinear terms in (\ref{eq:toypde2})
generates the coupling term in (\ref{eq:AT}), so if we set
$s=0$ there is no coupling and the dynamics is then governed by the
Ginzburg--Landau equation alone. Furthermore, if the original PDE
has the symmetry $w \to -w$ then all nonlinear terms in (\ref{eq:AToy2}) are
cubic and again no coupling terms arise.
This restriction will apply when we consider physical examples in
\S~\ref{sec:examples}.

\section{Analysis of the amplitude equations}
\label{sec:anal}

In this section we analyse the steady solutions
of the amplitude equations (\ref{eq:AT}--\ref{eq:BT}).
We begin by considering small-amplitude, periodic  solutions (rolls).
We then investigate the super/subcritical nature of the instability,
and, beyond the point where rolls become unstable, we calculate
fully nonlinear solutions of (\ref{eq:AT}--\ref{eq:BT}).

\subsection{Stability of patterns}
\label{sec:stabpat}

In this section we  analyse the
stability of periodic solutions in the form of rolls.
Rolls with wavenumber $k_c+\epsilon q$ correspond to solutions of
(\ref{eq:AT}--\ref{eq:BT}) of the form $A=a_0\e^{ \i q X}$ and $B=0$,
where the amplitude $a_0$ may be taken to be real and to satisfy $a_0^2=1-q^2$.
To analyse the stability of these rolls, we set
\beq
A=(a_0+a(X,T))\e^{ \i qX},
\qquad
B=b(X,T),
\label{eq:a0ab}
\eeq
and linearise (\ref{eq:AT}--\ref{eq:BT}) in $a$ and $b$.
Upon setting $a=V(T)\e^{\i lX}+W^*(T)\e^{-\i lX}$ and
$b=U(T)\e^{\i lX}+U^*(T)\e^{-\i lX}$, we find
\bea
\dot V & = & -(2q+l)lV-a_0^2(V+W)-a_0U ,\label{eq:V} \\
\dot W & = & -(-2q+l)lW-a_0^2(V+W)-a_0U , \label{eq:W}\\
\dot U & = & -\sigma l^2U-\mu a_0 l^2 (V+W).\label{eq:U}
\eea
Solutions of (\ref{eq:V}--\ref{eq:U}) proportional to $\e^{\lambda T}$
have a growth rate $\lambda$ that satisfies the cubic equation
\bea\fl
\lambda^3+(2l^2+\sigma l^2+2a_0^2)\lambda^2
 -l^2(2\mu a_0^2-2\sigma a_0^2-2a_0^2-l^2-2\sigma l^2+4q^2)\lambda \nonumber\\
 -l^4(2\mu a_0^2-2\sigma a_0^2-\sigma l^2+4\sigma q^2)  =  0.\label{eq:stabcub}
\eea
For modes with $\lambda$ real, there is instability if
\beq
2(\mu-\sigma)a_0^2+\sigma(4q^2-l^2)>0. \label{eq:crit0}
\eeq
The most dangerous disturbance therefore arises in the limit $l\to0$,
and so the criterion for instability is
\beq
(\mu-\sigma)a_0^2+2\sigma q^2>0,
\label{eq:crit1}
\eeq
or, equivalently,
\beq
\mu(1-q^2)>\sigma(1-3q^2).
\label{eq:crit2}
\eeq
From (\ref{eq:crit1}) it is clear that all rolls are unstable if
\beq
\mu>\sigma.    \label{eq:stabcond}
\eeq
If $\mu<\sigma$ then rolls of sufficiently small amplitude
are unstable;  the criterion for instability can be expressed in terms
of $q$ as
\beq
 q^2 > (\sigma - \mu)/(3 \sigma - \mu) . \label{eq:qstabcond}
\eeq
It is clear that when the coupling
between the mean and pattern modes is removed ($\mu=0$), or more generally
when $|\sigma/\mu|\gg1$, the usual Eckhaus
criterion $q^2 > 1/3$ for instability is recovered.
In general, the band of stable rolls is narrowed if $\mu > 0$ and
widened if $\mu < 0$.

Hopf bifurcations are also permitted in (\ref{eq:stabcub}).
We are able to make analytical progress in the
small-$l$ limit, where one solution of
(\ref{eq:stabcub}) is $\lambda = - 2 a_0^2$ while the other two obey
\beq
\lambda^2 a_0^2 + \lambda l^2 (a_0^2 (1+\sigma -\mu) - 2 q^2)
+ l^4 (\sigma - \mu - 3 q^2 \sigma + q^2 \mu) = 0 .
\eeq
It follows that there is oscillatory instability
if $q^2 > (1+\sigma-\mu)/(3+\sigma-\mu)$
provided that $\mu \sigma -\sigma^2 - \mu > 0 $.
From these conditions it can be deduced that a Hopf bifurcation is
possible only if $\sigma < 1$, $\mu < 0$ and $q^2 > 1/3$.
There is a Takens--Bogdanov point (a repeated zero eigenvalue)
at $q^2 = 1/(3 - 2\sigma)$ when $\mu (\sigma - 1) = \sigma^2$.

\begin{figure}
\begin{center}
\epsfxsize0.7\hsize\epsffile{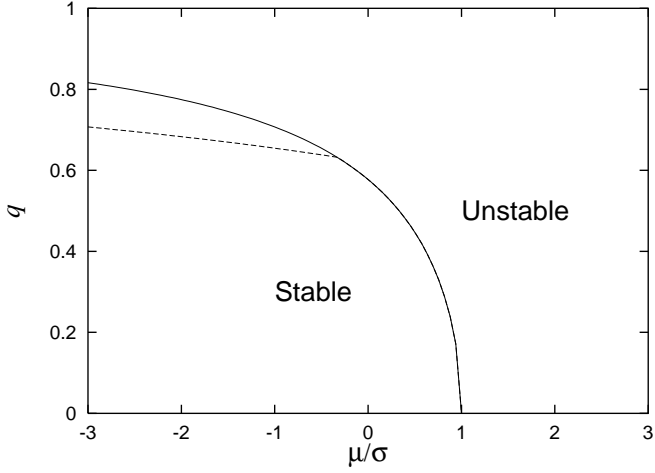}
\end{center}
\caption{Region of stable rolls in (\ref{eq:AT}--\ref{eq:BT}).
The solid line shows the stationary bifurcation, which is valid for
any $\sigma$,  and the dotted line
shows the Hopf bifurcation for $\sigma=1/4$.
For $\mu > \sigma$ all rolls are unstable, while for  $\mu < \sigma$
there is a band of stable rolls that widens as $\mu $ is decreased.
The usual Eckhaus result of stability for $q^2<\frac13$
corresponds to $\mu=0$. }
\label{fig:stab}
\end{figure}

The stability results are summarised in figure~\ref{fig:stab}.
Note that the stationary bifurcation given by (\ref{eq:qstabcond})
depends only on the ratio $\mu/\sigma$ but this is not true for the
Hopf  bifurcation.

The mechanism of instability in (\ref{eq:AT}--\ref{eq:BT}) is rather
different from that of the Eckhaus instability.
Even rolls with $q=0$ are unstable for $\mu > \sigma$, and in this case
the unstable mode has $V = W$ so that $a$ is real and the instability is
amplitude-driven rather than phase-driven.
If there are regions where $|A|^2$ has a minimum, then, according to
(\ref{eq:BT}), $B$ grows in these regions if $\mu > 0$.
This leads to a positive value of $B$ which then reduces
the size of $A$ further, through the coupling term in (\ref{eq:AT}).

\subsection{Nonlinear development of the instability}
\label{sec:ndi}

The results of \S~\ref{sec:stabpat} show that for $\mu > \sigma$,
all periodic patterns in (\ref{eq:AT}--\ref{eq:BT}) are unstable.
We now consider the nonlinear development of the instability,
to determine whether the bifurcation is subcritical or supercritical.
The Eckhaus instability, for example, is known to give rise to a
subcritical bifurcation (Fauve 1987, Tuckerman and Barkley 1990).

For simplicity we consider the case $q=0$, corresponding to
rolls with wavenumber $k_c$. These are the last rolls to become
unstable as $\mu$ is increased. In this case the stationary solution
is $A = 1$, $B = 0$.
Writing $A = 1 + a(X,T)$, $B = b(X,T)$, cf.\ (\ref{eq:a0ab}),
we find that the imaginary
part of $a$ decays so that we may take $a$ to be real.
The nonlinear equations for the real-valued quantities $a$ and $b$ are then,
from (\ref{eq:AT}--\ref{eq:BT}),
\bea
a_T = - 2 a + a_{XX} - 3 a^2 - a^3 - b - a b , \label{eq:3.3A}\\
b_T = \sigma b_{XX} + \mu (2 a + a^2 )_{XX} . \label{eq:3.3B}
\eea
A weakly nonlinear analysis
is now employed, by writing
\begin{eqnarray*}
\mu = \mu_0 + \ep^2 \mu_2, \qquad T_2 = \ep^2 T , \\
a = \ep a_1 + \ep^2 a_2 + O(\ep^3), \qquad
b = \ep b_1 + \ep^2 b_2 + O(\ep^3),
\end{eqnarray*}
where the $a_i$ and $b_i$ are functions of $X$ and $T_2$.
By considering a single mode,
\[
a_1 = a_{11}(T_2) \cos l X , \qquad b_1 = b_{11}(T_2) \cos l X,
\]
we find that the terms of order $\ep$ in (\ref{eq:3.3A}--\ref{eq:3.3B}) give
\[
b_{11} + (2+l^2) a_{11} = 0 \quad\mbox{and}\quad \mu_0 = (1 + l^2/2)\sigma ,
\]
so the instability condition (\ref{eq:stabcond}) is recovered for
small $l$.
At $O(\ep^2)$ we deduce that
\[
a_2 = a_{20} + a_{22}\cos 2 l X, \qquad b_2 = b_{22}\cos 2 l X,
\]
where
\[
a_{20} = (l^2 -1)a_{11}^2/4, \quad a_{22} = a_{11}^2/4
\quad\mbox{and}\quad b_{22} = - (1+l^2/2) a_{11}^2 .
\]
Applying the solvability condition at $O(\ep^3)$
gives the equation for the evolution on the slow timescale $T_2$:
\[
(\sigma l^2 + 2 + l^2) \d {a_{11}} {T_2} = 2 l^2 \mu_2 a_{11}
+ \frac {\sigma l^4(l^2 -1 )} {2}  a_{11}^3.
\]
Hence the bifurcation is a supercritical pitchfork provided that
$l < 1$.
In a large periodic domain, the first modes to become unstable have small $l$,
and so the bifurcation will in practice be supercritical.

\subsection{Nonuniform steady solutions}\label{sec:nolinsol}

We now extend the analysis of the previous subsection to consider
fully nonlinear stationary solutions of the amplitude equations
(\ref{eq:AT})--(\ref{eq:BT}). Writing $A=R\exp\i\Theta$, we find from
the imaginary part of (\ref{eq:AT}) that $(R^2\Theta_X)_X=0$ and so 
$\Theta_X=\Omega R^{-2}$, where $\Omega$ is constant.
From (\ref{eq:BT}) we find, for solutions periodic in $X$,
that
\beq
B=\mu'(\langle R^2\rangle -R^2),
\label{eq:arghhh}
\eeq
where $\langle\cdots\rangle$ denotes a spatial average. In
deriving (\ref{eq:arghhh}), we have taken $\langle B\rangle=0$,
which corresponds to an initial condition such that $\langle
w\rangle=0$ (and hence $\langle w\rangle=0$ for all time). Any
other initial value for $\langle w\rangle$ is equivalent to
$\langle w\rangle=0$ under a rescaling of the parameters $r$ and
$s$ in (\ref{eq:toypde2}). 
From the real part of (\ref{eq:AT}), it then follows that $R$ satisfies
\beq
0=(1-\mu'\langle R^2\rangle)R+R_{XX}-(1-\mu')R^3-\Omega^2R^{-3}.
\label{eq:R}
\eeq
This equation may be integrated once, and is then more conveniently 
expressed in terms of $Q=R^2$:
\beq
{Q'}^2=2(1-\mu')Q^3+4(\mu'\langle Q\rangle -1)Q^2+4HQ-4\Omega^2,
\eeq
where $H$ is a constant of integration.  
The solution is readily determined in terms of the Weierstrass elliptic
function ${\cal P}$ (Abramowitz and Stegun 1965, Whittaker and Watson 1962)
to be
\beq
Q=\frac{2}{(1-\mu')}{\cal P}(X;g_2,g_3)+{\cal B},
\label{eq:exactQ}
\eeq
where
\bea
g_2&=&- 2(1-\mu')H+\tsfrac{4}{3}(\mu'\langle Q\rangle-1)^2 \label{eq:g2}\\
g_3&=&
\tsfrac{2}{3}(1-\mu') (\mu'\langle Q\rangle -1) H
-\tsfrac{8}{27}(\mu'\langle Q\rangle -1)^3
+(1-\mu')^2 \Omega^2
                   \\
{\cal B}&=&\frac{-2(\mu'\langle Q\rangle-1)}{3(1-\mu')},\label{eq:calB}
\eea
and $g_2$ and $g_3$ are the usual parameters of ${\cal P}$.
The amplitude and period of the solution are implicitly related through
the appearance of the quantity $\langle Q\rangle$ in  
(\ref{eq:g2})--(\ref{eq:calB}).

Although (\ref{eq:exactQ}) provides an exact solution, it requires
specification of $\mu'$, $H$, $\Omega$ and the spatial period to permit
analysis of its solutions.
Rather more analytical progress is possible if we 
restrict attention to the invariant
subspace in which $A$ is real. We suppose that the solution has
spatial period $L$ (note that, as a consequence, $\epsilon
L=2n\pi$ for some integer $n$). Equations
(\ref{eq:AT})--(\ref{eq:BT}) become
\bea
0 & = & A + A_{XX} - A^3 - AB  ,           \label{eq:AS}\\
0 & = & B_{XX} + \mu' (A^2)_{XX},           \label{eq:BS}
\eea
where
\[
\mu'=\mu/\sigma.
\]
Integrating (\ref{eq:BS}) twice and imposing periodicity gives the
result
\beq
B + \mu' A^2 = \mu'\langle A^2 \rangle,
\eeq
corresponding to (\ref{eq:arghhh}).
Eliminating $B$ from (\ref{eq:AS}) gives
\beq
0 = (1-\mu'\langle A^2 \rangle)A + A_{XX} - (1 - \mu') A^3 . \label{eq:aode}
\eeq

Before describing the periodic solutions of
(\ref{eq:AS})--(\ref{eq:BS}), it is instructive to consider a
solution that is a good asymptotic approximation:
\beq A = A_0
\,\mbox{sech} (c X), \label{eq:sechsol}
\eeq
where $\mu'\langle A^2\rangle = 1 + c^2$ and $\mu' = 1 + 2 c^2 /
A_0^2$. Note that the solution exists only for $\mu' > 1$, which
is consistent with the stability boundary (\ref{eq:stabcond}) and
the result from \S~\ref{sec:ndi} that the bifurcation at $\mu' =
1$ is supercritical. For large values of $c L$, $cL\langle
A^2\rangle \sim2 A_0^2$. Using this approximation for $\langle
A\rangle$, we obtain for $c$ the quadratic equation
\[
c^2 - 4 \mu' c / L(\mu' -1) + 1 = 0 ,
\]
which has real solutions only for
\beq
1 < \mu' < L/(L-2) , \label{eq:sn}
\eeq
indicating that a saddle--node bifurcation occurs  at $\mu' =
L/(L-2)$ and that for $\mu' >L/(L-2)$ no such solution exists. At
large amplitude the solution branch approaches $\mu'=1$ and the
amplitude is given by $A_0 \sim 4\sqrt 2 \, \mu' / L
(\mu'-1)^{3/2}$. The solution (\ref{eq:sechsol}) is of particular
significance because, as we shall see, it can be stable and is
observed in the numerical simulations described in
\S~\ref{sec:numsims}.


Spatially periodic solutions of (\ref{eq:aode}) can be written in terms
of the Jacobi elliptic functions $\Jsn$, $\Jcn$ and $\Jdn$. In all that
follows, $m$ denotes the standard parameter of these functions
(Whittaker and Watson 1962, Abramowitz and Stegun 1965),
and we use the notation $\langle\cdots\rangle$ as above, so that
\[
\langle f(X) \rangle\equiv \frac{1}{4K} \int_0^{4K} f(X)\,{\rm d}X,
\]
where $K$ is the real quarter-period, given by
\[
K(m)=\int_0^{\pi/2}\frac{{\rm d}\theta}{(1-m\sin^2\theta)^{1/2}}.
\]

\subsubsection{dn}\label{sec:dn}

When $\mu'>1$, there exist solutions of (\ref{eq:aode})
in the form
\beq
A=A_0 \Jdn \, c X,
\label{eq:Adn}
\eeq
where
\beq\fl
(2-m) c^2=A_0^2\mu'\langle\Jdn^2 cX\rangle-1,
\qquad 2 c^2=A_0^2(\mu'-1),\qquad
 c L=2K.
\label{eq:dn12}
\eeq
Given the parameter $\mu'$ and the length $L$ of the domain, the three
equations in (\ref{eq:dn12}) serve to determine the three
unknowns $A_0$, $c$ and $m$. However, the computation of
solutions is more easily accomplished if $L$, say, is fixed, while $m$
is varied in the range $0\leq m<1$ to yield corresponding values of
$A_0$, $c$ and $\mu'$.

For small $m$, the solution takes the form
\beq
A\sim(1+\tsfrac14 m)(1-\tsfrac12 m\sin^2(\pi X/L)),
\eeq
indicating that these solutions branch from the uniform state $A\equiv1$;
the branching takes place at $\mu'=1+2\pi^2/L^2$ (corresponding to $m=0$).
Furthermore, it is readily shown from (\ref{eq:dn12}) that for small $m$
\beq
\mu'=1+\frac{2\pi^2}{L^2}+\frac{\pi^2}{16L^4}(L^2-4\pi^2)m^2+O(m^3),
\eeq
confirming the result of \S~\ref{sec:ndi} that the bifurcation is
supercritical when $L>2\pi$ ($l<1$ in the notation of  \S~\ref{sec:ndi})
and subcritical when $L<2\pi$ ($l>1$).

\begin{figure}
\epsfxsize0.7\hsize\epsffile{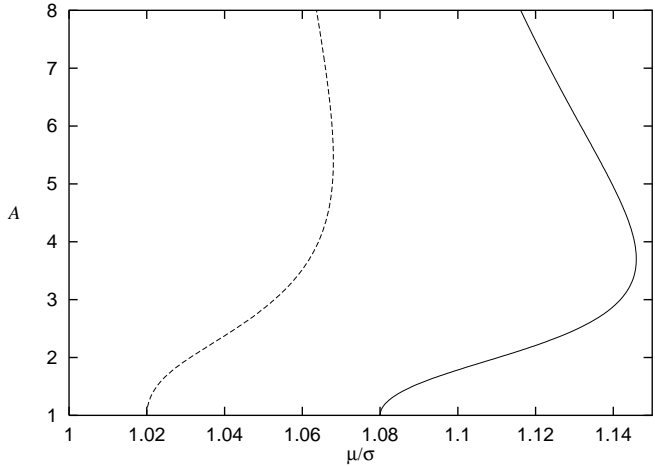}
\caption{The bifurcation diagram for (\ref{eq:AS})--(\ref{eq:BS}) with periodic
boundary conditions with $L=5\pi$ (solid line) and $L=10\pi$ (dashed
line). Solutions take the form (\ref{eq:Adn}).
The vertical axis shows the maximum value of $A$.\label{fig:bif1}}
\end{figure}

As $m$ is increased from zero to unity, the corresponding
values of $\mu'$ increase to a maximum $\mu_d$ (given approximately
in (\ref{eq:sn}) by $\mu_d\sim L/(L-2)$),
and then decrease monotonically so that
\beq
\mu'\sim 1+\frac{4}{\ln(16/m_1)},
\label{eq:mum1-}
\eeq
as $m\to1^-$, where $m_1=1-m$. We have used the well known result that
\beq
K\sim\ln(4/m_1^{1/2})
\label{eq:Klim}
\eeq
in this limit.
Thus a saddle--node bifurcation takes place at $\mu'=\mu_d$,
and for $\mu'>\mu_d$ no solutions of the form (\ref{eq:Adn}) exist.
Figure~\ref{fig:bif1} shows the bifurcation diagram
for dn-type solutions of (\ref{eq:AS})--(\ref{eq:BS}) subject to periodic
boundary conditions with $L = 5 \pi$ and $L = 10 \pi$.
The saddle--node bifurcations occur at $\mu_d=1.146$ and $\mu_d=1.068$
respectively, in agreement with our approximation (\ref{eq:sn}).
Since the bifurcation is supercritical in each case,
the lower branch of solutions is stable but the upper branch is unstable.

\begin{figure}
\centerline{
\epsfxsize0.5\hsize\epsffile{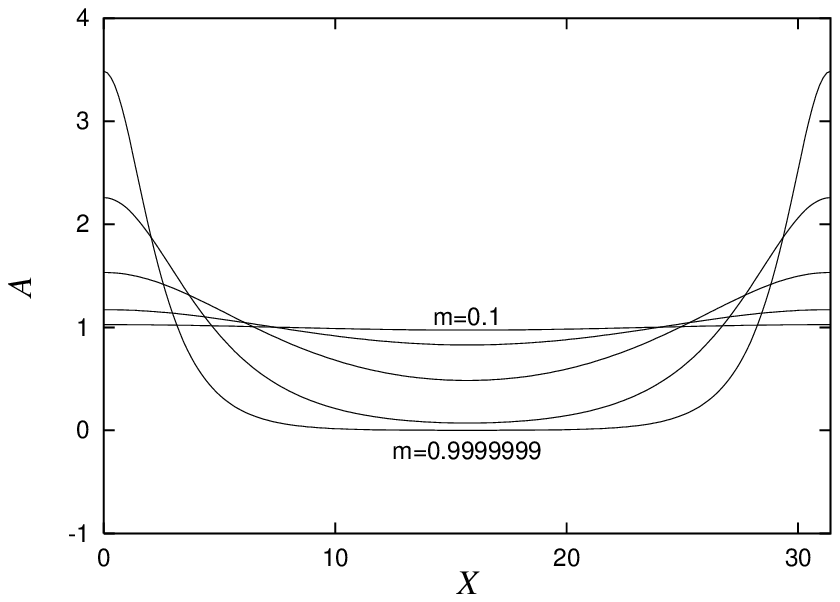}
\epsfxsize0.5\hsize\epsffile{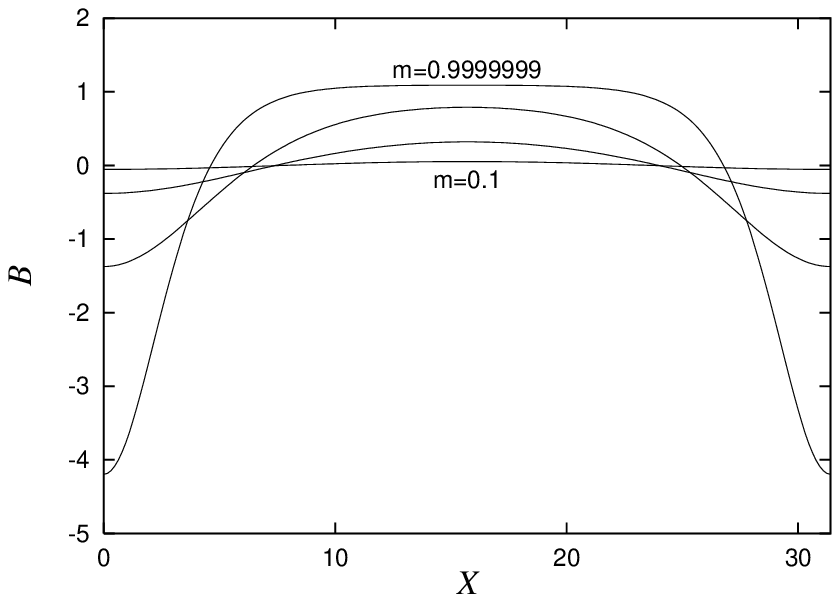}}
\caption{Solution of (\ref{eq:AS}) and (\ref{eq:BS}) in terms of dn
(\ref{eq:Adn}) for $m=0.1$, $0.5$, $0.9$, $0.999$, $0.9999999$
(corresponding to
$\mu'=1.0200$, $1.0203$, $1.0230$, $1.0504$, $1.0596$).
The spatial period is $L=31.4$.
\label{fig:dn}}
\end{figure}

When $m$ is small, the solution for $A$ takes the form of small,
almost sinusoidal perturbations to the uniform state $A\equiv1$. As $m$ is
increased, the peak of $A$ becomes narrow and the trough broadens. As
$m\to1^-$, the peak increases greatly in amplitude and the solution takes
the form
\beq
A\sim \frac{2K^{3/2}}{L}\sech\frac{2KX}{L}
\label{eq:dnsech}
\eeq
where $\mu'\sim1+2/K$, and $K\to\infty$ according to (\ref{eq:Klim}). In
this limit, the solution for $A$ thus takes the form of a single sech-type
pulse as given by (\ref{eq:sechsol}).
The dn solutions are illustrated in figure~\ref{fig:dn}.

These results indicate that for $\mu_d>\mu' > 1$ a soliton-like
solution exists to (\ref{eq:AT})--(\ref{eq:BT}).
For $\mu'>\mu_d$, numerical simulations of (\ref{eq:AT})--(\ref{eq:BT})
indicate that the solution blows up, signalling a transition to a
fully nonlinear solution of the original PDEs (for example (\ref{eq:toypde2})).
This blow-up can be avoided by adding higher-order terms to the
amplitude equations, which we discuss in \S~\ref{sec:hot}.

\subsubsection{cn}

When $\mu'>1$, solutions also exist in the form
\beq
A=A_0 \Jcn \, c X,
\label{eq:Acn}
\eeq
where
\beq\fl
 (2m-1) c^2=A_0^2\mu'\langle\Jcn^2 cX\rangle-1,
\qquad 2m c^2=A_0^2(\mu'-1),\qquad
c L=4K.
\label{eq:cn12}
\eeq

\begin{figure}
\epsfxsize0.6\hsize\epsffile{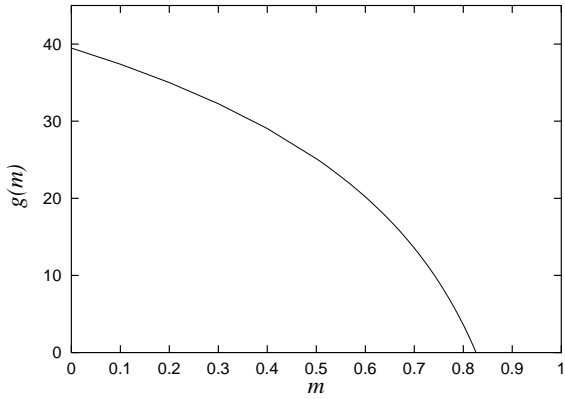}
\caption{The function $g(m)$, defined in (\ref{eq:g(m)}). The intercepts with
the axes are given by $g(0)=4\pi^2$ and $g(m_0)=0$, where $m_0\approx0.826115$.
\label{fig:g(m)}}
\end{figure}

When $m$ is small, $A\sim \sqrt{2}(1-4\pi^2/L^2)^{1/2}\cos(2\pi X/L)$, and
hence we require $L>2\pi$ for the existence of this solution in this limit.
Further analysis indicates that for general values of $m$ a solution exists
only when
\beq
L^2>g(m)\equiv 16K^2[1+2m(\langle \Jcn^2 c X \rangle-1)].
\label{eq:g(m)}
\eeq
The function $g(m)$ is shown in figure~\ref{fig:g(m)}. It is clear from
(\ref{eq:g(m)}) and figure~\ref{fig:g(m)} that if $L>2\pi$ the solution
(\ref{eq:Acn}) exists for all $m$ between zero and unity. If, on the other
hand, $L<2\pi$ then solutions exist for $m_L<m<1$, where $g(m_L)=L^2$.
Regardless of the value of $L$, these solutions exist for $m>m_0\approx
0.826115$.
If $L>2\pi$ then
as $m$ is increased from zero, $\mu'$ increases from unity to a maximum
value $\mu_c$ (given approximately by $\mu_c=L/(L-4)$)
and then decreases monotonically according to (\ref{eq:mum1-}) as $m\to1^-$.
If $L<2\pi$ then as $m$ is increased from $m_L$, $\mu'$ decreases
monotonically from infinity,
again according to (\ref{eq:mum1-}) as $m\to1^-$.

\begin{figure}
\centerline{
\epsfxsize0.5\hsize\epsffile{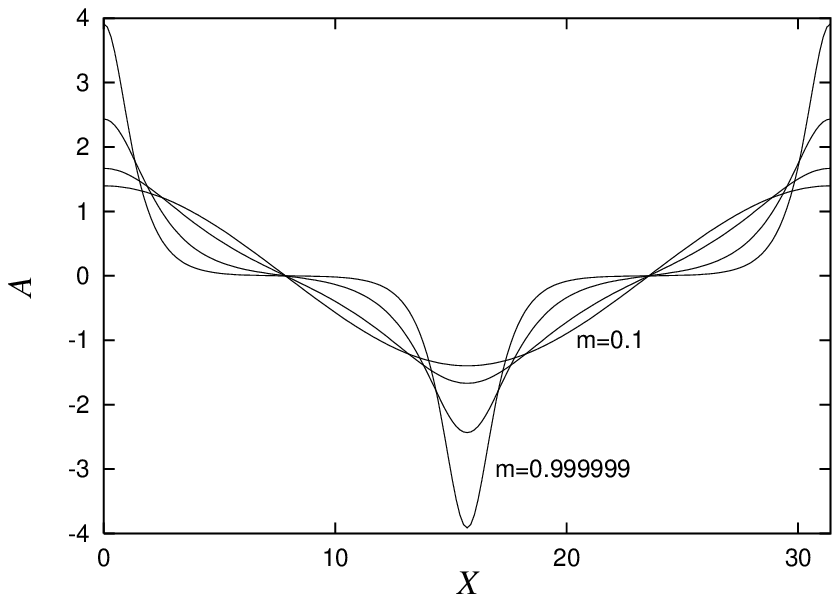}
\epsfxsize0.5\hsize\epsffile{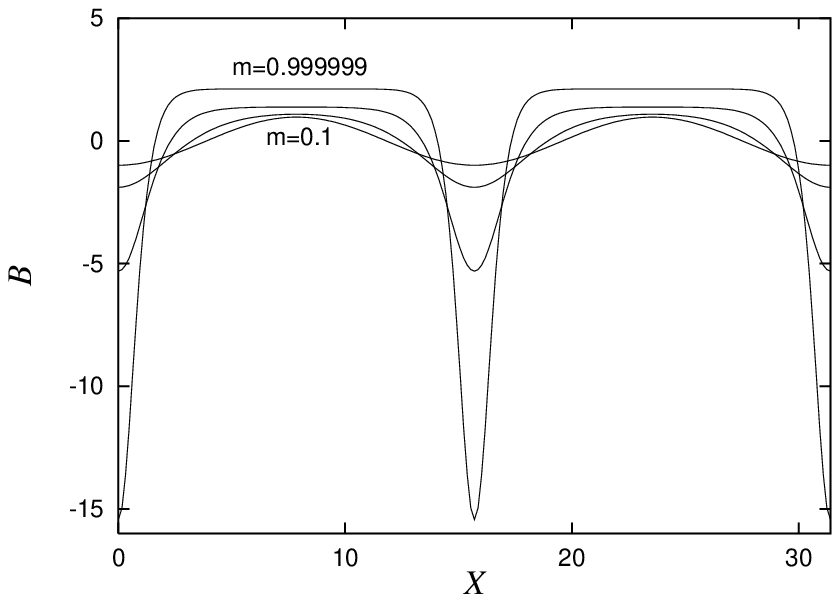}}
\caption{Solution of (\ref{eq:AS}) and (\ref{eq:BS}) in terms of cn
(\ref{eq:Acn}) for $m=0.1$, $0.9$, $0.999$, $0.999999$ (corresponding to
$\mu'=1.004$, $1.069$, $1.128$, $1.146$).
The spatial period is $L=31.4$.
\label{fig:cn}}
\end{figure}

As $m$ is
increased, the peaks of $A$ become more pronounced, larger in amplitude
and narrower, while the regions in which $A$ has small amplitude become
broader. In the limit $m\to1^-$, the solution is approximately given
by two sech-type pulses, one positive and one negative; with an
appropriate choice of origin in $X$, the solution for $0<X<L$ satisfies
\[
A\sim\frac{4K^{3/2}}{L}\left[\sech \frac{4KX}{L}-
\sech\frac{4K(X-L/2)}{L}\right],
\]
with $K$ and $m$ related through (\ref{eq:Klim}).
The cn solutions are illustrated in figure~\ref{fig:cn}.

\subsubsection{sn}

The final type of solution that may be expressed in the form of Jacobi
elliptic functions applies when $\mu'<1$, and is
\beq
A=A_0 \Jsn \, c X,
\label{eq:Asn}
\eeq
where
\beq\fl
(1+m) c^2=1-A_0^2\mu'\langle\Jsn^2 cX\rangle,
\qquad 2m c^2=A_0^2(1-\mu'),\qquad
 c L=4K.
\label{eq:sn12}
\eeq

\begin{figure}
\centerline{
\epsfxsize0.5\hsize\epsffile{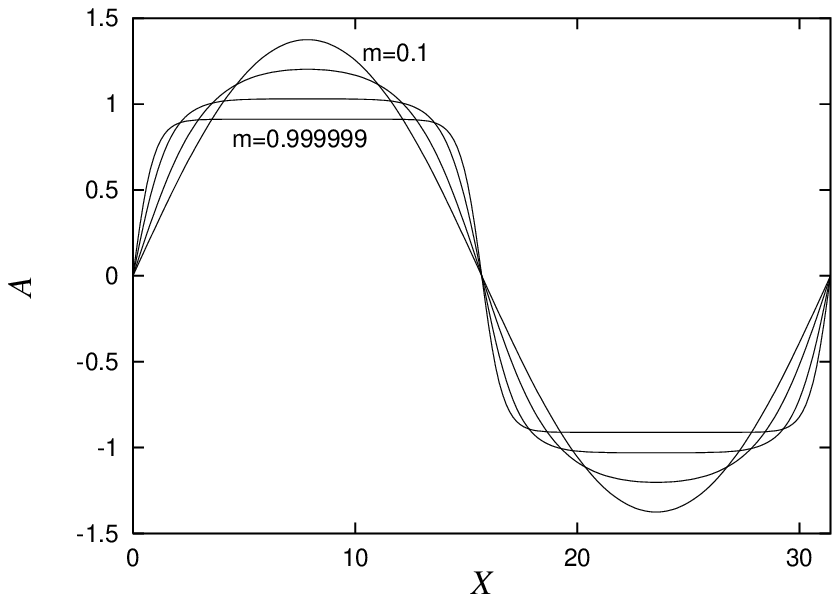}
\epsfxsize0.5\hsize\epsffile{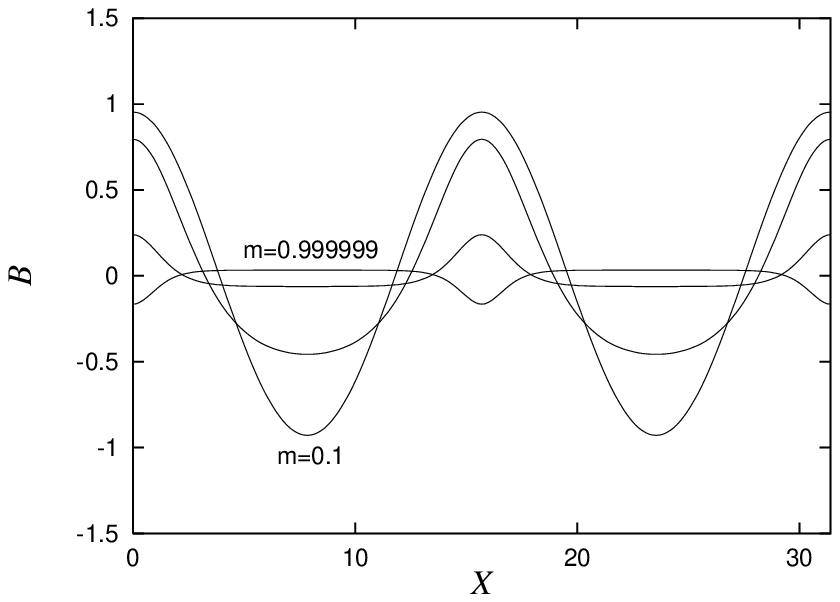}}
\caption{Solution of (\ref{eq:AS}) and (\ref{eq:BS}) in terms of sn
(\ref{eq:Asn}). The first plot shows $A$ for $m=0.1$, $0.9$,
$0.999$, $0.999999$
(corresponding to $\mu'=0.9955$, $0.8658$, $0.2844$, $-1.6867$).
The second plot
shows $B$ for these values of $m$ and in addition $m=0.9999$ and $0.99999$
($\mu'=-0.2049$, $-0.8538$).
The spatial period is $L=31.4$.
\label{fig:sn}}
\end{figure}

For small $m$, these solutions match with the cn solutions (allowing for
an unimportant translation in $X$), with
$A\sim \sqrt{2}(1-4\pi^2/L^2)^{1/2}\sin(2\pi X/L)$.
As this expression indicates, the small-$m$ solutions require $L>2\pi$.
In fact, for general values of $m$, the length $L$ of the domain must satisfy
\beq
L^2>h(m)\equiv 16K^2[1+m-2m\langle \Jsn^2 c X \rangle].
\label{eq:h(m)}
\eeq
The function $h(m)$ is monotonic increasing, with $h(0)=4\pi^2$ and
$h(m)\to\infty$ as $m\to1^-$. Thus solutions of the form (\ref{eq:Asn})
do not exist if $L<2\pi$; if $L>2\pi$, solutions exist for
$0<m<m_L$, where $h(m_L)=L^2$.
As $m$ increases towards $m_L$, $\mu'$ decreases monotonically to $-\infty$,
the peaks and troughs in $A$ becoming broader. Provided $L$ is
large, the profile near $X=0$ is approximately a front of the form
\[
A\sim\tanh\frac{4KX}{L}
\]
as $m\to m_L^-$,
the front near $X=L/2$ being given by symmetry.
The sn solutions are illustrated in figure~\ref{fig:sn}.

\begin{figure}
\epsfxsize0.7\hsize\epsffile{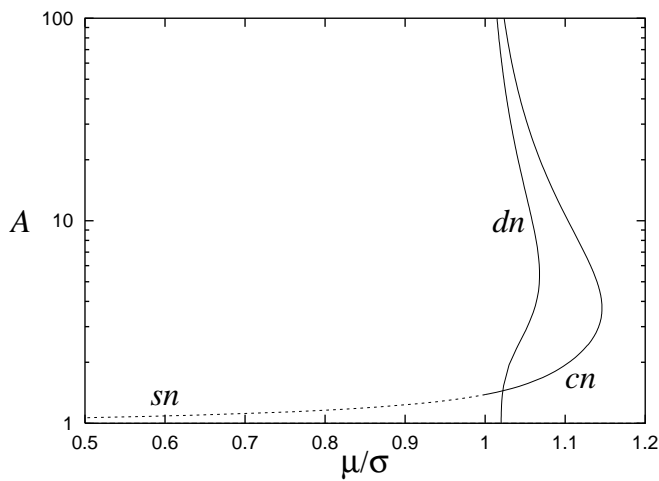}
\caption{The bifurcation diagram for (\ref{eq:AS})--(\ref{eq:BS}) with periodic
boundary conditions with $L=10\pi$, showing dn, cn (both solid lines) and
sn (dashed line) solution branches.
The vertical axis shows the maximum value of $A$.\label{fig:bifcsd}}
\end{figure}

A bifurcation diagram showing all three types of solution (cn, dn and
sn) is given in figure~\ref{fig:bifcsd}.
Note that only the dn solutions bifurcate from the periodic rolls
represented by $A=1$.

The solutions described above are those containing the minimal number of
peaks and troughs within the periodic domain. Other solutions,
containing $n$ sets of peaks and troughs can also be
constructed by replacing $K$ by $nK$ in the last equation of
(\ref{eq:dn12}), (\ref{eq:cn12}) or (\ref{eq:sn12}).

\section{Numerical simulations of a model PDE}
\label{sec:numsims}

In this section we illustrate the application of the results of the
previous sections through numerical simulations of the PDE
(\ref{eq:toypde2}), assuming periodic boundary conditions.
From the analysis of  \S~\ref{sec:toypde}, 
the system is governed by the amplitude equations
(\ref{eq:AT})--(\ref{eq:BT}) with
\beq
\sigma = \frac 1 4 \qquad\mbox{and}\qquad \mu = \frac{s^2}{3-2s^2/9}.
\label{eq:sigmu}
\eeq
A pseudospectral method is employed in which the nonlinear terms are
computed in physical space using FFTs. 
Time advancement is carried out by approximating the nonlinear terms as
a linear function of time and then obtaining exact solutions
using the  known growth rate of the linear terms  (\ref{eq:lambda}).

\begin{figure}
(a)\\
\epsfxsize0.9\hsize\epsffile{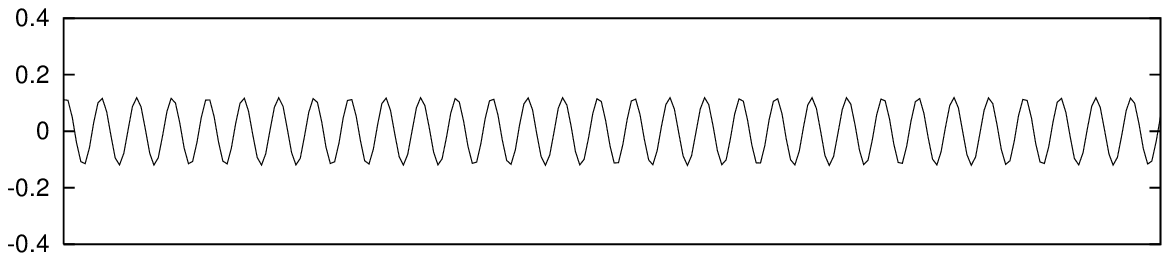}\\
(b)\\
\epsfxsize0.9\hsize\epsffile{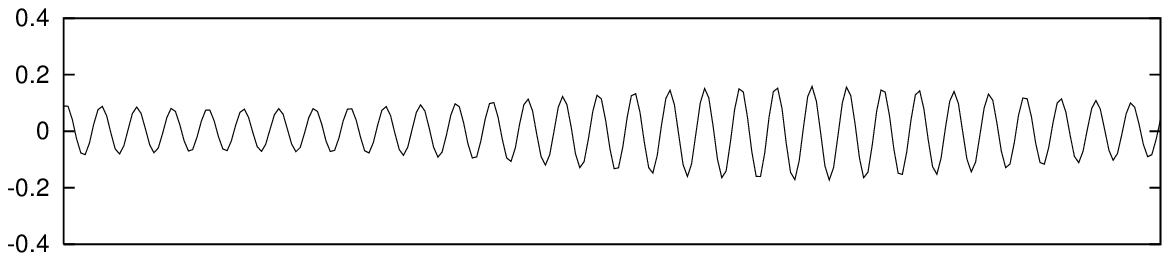}\\
(c)\\
\epsfxsize0.9\hsize\epsffile{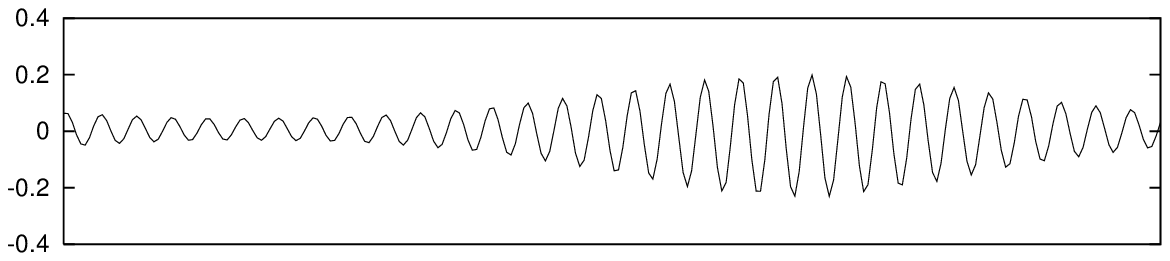}\\
(d)\\
\epsfxsize0.9\hsize\epsffile{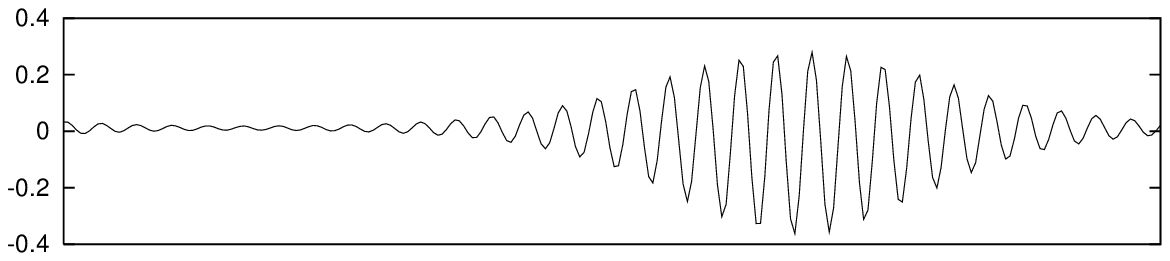}
\caption{Numerical simulations of (\ref{eq:toypde2}) showing $w(x)$,
with $r=0.01$ and periodic domain size $62\pi$.
(a) Stable rolls, $s=0.9$; (b) Modulated rolls, $s=0.93$;
(c) $s=0.95$; (d) $s=1.0$.
\label{fig:numsims}}
\end{figure}

In order to approach the asymptotic regime, the size of the periodic
domain must be large and the driving parameter $r$ must be small.
Initially, the length of the domain is chosen to be $62 \pi$, so
that $31$ periods of the pattern are contained in the box,
the forcing parameter is
$r = 0.01$ and the number of grid points in physical space is 256.
The effective length of the domain for the amplitude equations
(\ref{eq:AT})--(\ref{eq:BT}) in terms of the lengthscale $X$ is then
$L=62 \pi \sqrt{r} / 2 = 3.1\pi$. Note that since $L>2\pi$,
the result of \S~\ref{sec:ndi} indicates that the bifurcation
from a regular pattern to a modulated pattern is supercritical.
In the limit of large $L$ the condition (\ref{eq:stabcond}) for rolls
to be unstable is $\mu > \sigma$, which is equivalent to
$s^2 > 27/38$. For finite $L$ the condition is (\ref{eq:crit0}), which yields
$\mu > \sigma(1 + 2\pi^2/L^2)$, or $|s| > 0.9215$ for $L=3.1\pi$.
In our numerical simulations of (\ref{eq:toypde2}) at $r=0.01$, we find 
good agreement with this asymptotic result,
rolls being unstable for $|s| > 0.927$.

A sequence of numerical solutions of (\ref{eq:toypde2}) with $r=0.01$ for
different values of $s$  is shown in figure~\ref{fig:numsims}. In
all cases, the solution shown is the final steady state after all
transients have decayed; to reach this state the computation must
be continued for several thousand time units. In
figure~\ref{fig:numsims}(a),  $s=0.9$ and a stable periodic
solution is found. For $s=0.93$ (figure~\ref{fig:numsims}(b)),
slightly beyond the bifurcation, a modulated pattern appears. The
amplitude of the modulation is small, which is consistent with the
result of \S~\ref{sec:ndi} that the bifurcation is supercritical.
As $s$ is increased, the amplitude of the modulation steadily
increases (figure~\ref{fig:numsims}(c,d)), until there are regions
in which the pattern is almost completely suppressed and regions
in which  the amplitude of the pattern becomes large. The envelope
of the pattern takes the form of the $\mbox{dn}$ or $\mbox{sech}$
solutions described in \S~\ref{sec:nolinsol} (see
figure~\ref{fig:dn});  we have not found any of the cn or sn
solutions, which suggests that they are unstable.

The increasing amplitude of these
solutions means that the numerical simulations are no longer within
the asymptotic regime of validity of the equations
(\ref{eq:AT})--(\ref{eq:BT}); in particular, for this value of $r$ 
we do not find the saddle--node bifurcation, which,
according to (\ref{eq:sn}), should occur when $s=0.939$ in
(\ref{eq:toypde2}). 

\begin{figure}
\epsfxsize0.9\hsize\epsffile{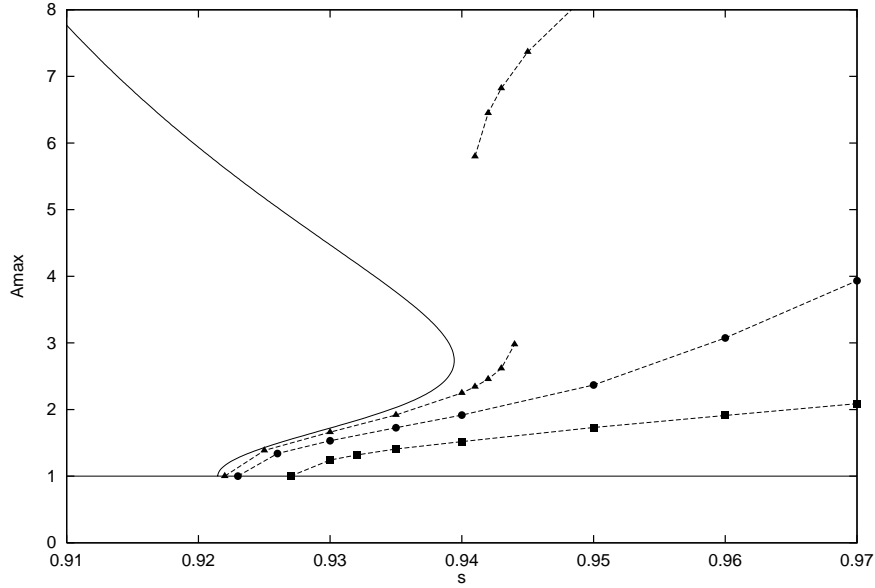}
\caption{Comparison of the solution to the amplitude equations
(\ref{eq:AT})--(\ref{eq:BT}) with $L = 3.1\pi$ (solid line) 
and numerical solutions for the same value of $L$ with three different
values of $r$.  Squares: $r=0.01$; circles: $r=0.0025$; triangles:
$r=0.000625$.
\label{fig:numbif}}
\end{figure}

To find this saddle--node bifurcation it is necessary to carry out
further sequences of numerical simulations at smaller values of $r$.
In order to keep the value of $L$ constant, the length of the domain
(and the number of grid points) must be increased in proportion to
$r^{-1/2}$.  The numerically obtained bifurcation diagrams, with $r$
successively reduced by a factor of 4,  are 
compared with the solution  of the amplitude equations in
figure~\ref{fig:numbif}.  The function $A$ is obtained numerically
from the part of the spectrum of the numerical solution near $k=1$ and
scaled appropriately with $r^{1/2}$.  For a domain size $62\pi$ with
$r=0.01$ there is no saddle--node bifurcation and the numerical and
theoretical bifurcation diagrams are not in close agreement. With a
domain size $124\pi$ and $r=0.0025$, agreement is better near the
pitchfork bifurcation but there is still no saddle--node
bifurcation. Finally, for a domain size $248\pi$ with $r=0.000625$ a
saddle--node bifurcation appears at $s = 0.944$, close to the
theoretical value of 0.939. Beyond this point the numerical results
jump to a solution of much larger amplitude, and there is a region of
hysteresis where two stable solutions exist.

\begin{figure}
(a)\\
\epsfxsize0.9\hsize\epsffile{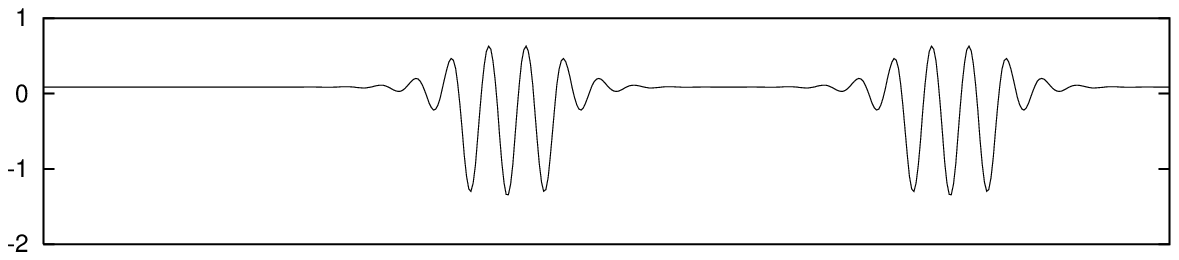}\\
(b)\\
\epsfxsize0.9\hsize\epsffile{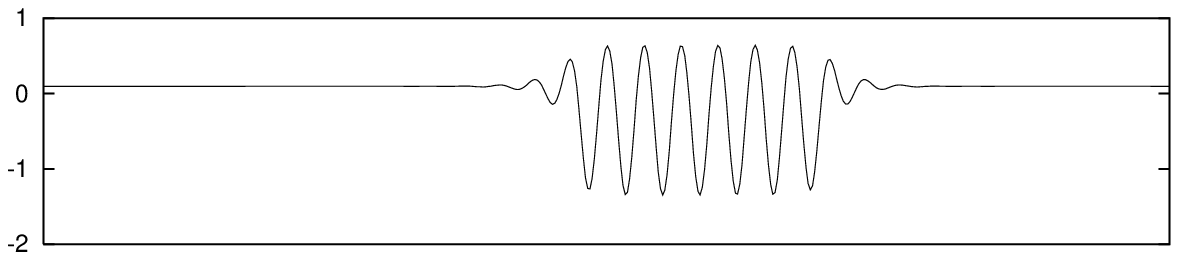}\\
\caption{Numerical simulations of (\ref{eq:toypde2}) showing $w(x)$,
with $r=0.01$, domain size $62\pi$ (using 512 grid points) and $s=1.6$.
The two solutions are started from different initial conditions.
The single-pulse solution appears to be a steady state, while in (a)
the two pulses are slowly separating. The amplitude appears
to be almost uniform in each of the humps.
\label{fig:numsims2}}
\end{figure}

For larger values of $s$, finite-amplitude solutions
of (\ref{eq:toypde2}) are found, that cannot be described in terms of
the amplitude equations. In some cases there are multiple stable
solutions.
For example, in figure~\ref{fig:numsims2} we show two solutions
computed at $s=1.6$, from different initial states.
In one plot there are two regions of pattern, and in the other only one.
The amplitude of the solution appears to be approximately the same in
each simulation.
However, the single pulse has roughly twice the width of each of the smaller
pulses, so in each simulation an equal proportion of the computation
domain seems to be occupied by the pattern.
Furthermore, it is clear that the envelope is rather broader than a
$\mbox{sech}$ profile, particularly in figure~\ref{fig:numsims2}(b), where
the envelope consists of two nearly uniform regions separated by a pair
of fronts.
In addition to those displayed in figure~\ref{fig:numsims2}, we
have found further steady (or quasi-steady) solutions of (\ref{eq:toypde2}),
including a single-pulse solution containing six large waves (rather
than the seven in figure~\ref{fig:numsims2}(b)) and a two-pulse solution
where the pulses are of unequal size, one containing three, and the other
four large waves.
Closer examination of the numerical results indicates that the multi-pulse
solutions are not quite stationary; the pulses are gradually moving.

\section{Beyond the leading-order amplitude equations}
\label{sec:hot}

In the previous section we have seen that 
agreement between  the amplitude equations
(\ref{eq:AT})--(\ref{eq:BT}) and the behaviour of the full PDE
(\ref{eq:toypde2})  only occurs for extremely small values of $r$. 
This agreement can be improved by introducing higher-order terms
into (\ref{eq:AT})--(\ref{eq:BT}).
Rather than include all possible terms that might arise at the next order
in (\ref{eq:AT})--(\ref{eq:BT}), we focus on one term.

To motivate our choice of higher-order term, we note that
when we compute the $x$-average of $w^3$, the leading-order
contribution is $\epsilon^4(\frac23s|A|^4+6|A|^2B)$. For moderate values
of $s$, the term $6\epsilon^4|A|^2B$ dominates. After rescaling the
equations for $A$ and $B$ so that $A$ satisfies (\ref{eq:AT}), we
find that $B$ satisfies
\beq
B_T  =  \sigma B_{XX} + \mu (|A|^2)_{XX}+\delta(B|A|^2)_{XX},
\eeq
where $\sigma$ and $\mu$ are given by (\ref{eq:sigmu}) and
\beq
\delta=\frac{3\epsilon^2}{2\left(3-\tsfrac29s^2\right)}>0.
\eeq
For future convenience we introduce the quantity $M$, defined by
\beq
\langle(M-\mu'|A|^2)/(1+\delta|A|^2/\sigma)\rangle=0.
\label{eq:Mint}
\eeq
When $\mu'>0$, the term in $\delta$ plays a stabilising role.

\begin{figure}
\epsfxsize0.9\hsize\epsffile{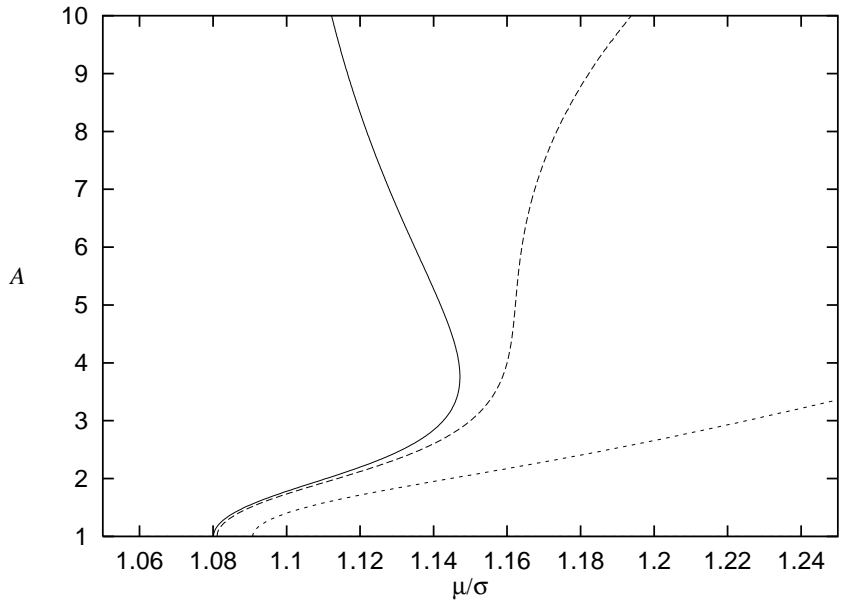}
\caption{Bifurcation diagram for (\ref{eq:Adel}) subject to (\ref{eq:Mint})
for $\delta/\sigma=10^{-4}$ (solid line), $10^{-3}$
(dashed line) and $10^{-2}$ (dotted line) plotted using AUTO97.
The domain length  is $L=5\pi$.
Note that very small values of $\delta$ can change
the qualitative nature of the bifurcation diagram, for example
eliminating the saddle--node bifurcation at $\delta=0$.
\label{fig:delbifs}}
\end{figure}

For steady solutions, the governing equation for $A$ becomes
\beq
0=A+A_{XX}-A|A|^2-A\left(\frac{M-\mu'|A|^2}{1+\delta|A|^2/\sigma}\right).
\label{eq:Adel}
\eeq
The bifurcation diagram for this equation, subject to the constraint
(\ref{eq:Mint}), is indicated in figure~\ref{fig:delbifs}. The bifurcation
diagram is changed significantly even by very small values of $\delta$.
For example, whereas there is a single saddle--node bifurcation for
$\delta=0$, when $\delta>0$ there is a pair of saddle--node bifurcations.
This pair exists only for $\delta/\sigma$ less than approximately
0.0009, and for larger values of $\delta/\sigma$ the curve is monotonic.

\begin{figure}
\epsfxsize0.9\hsize\epsffile{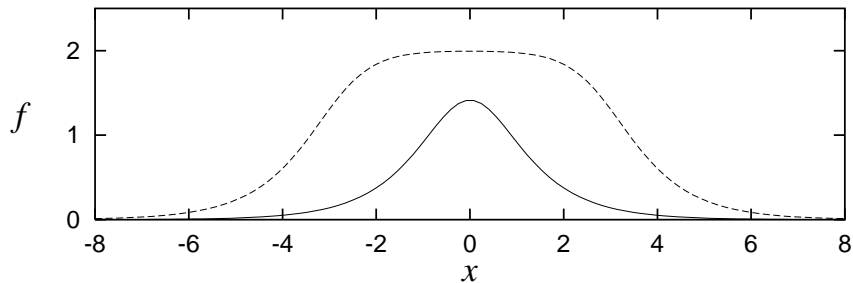}
\caption{Envelopes (\ref{eq:fsol}) from the model equation (\ref{eq:fmod}) with
$\delta=0$ (solid line) and $\delta=0.185$ (dashed line). The larger
value of $\delta$ produces a broad profile with a nearly uniform region
in which $f$ is non-zero. The broader profile is reminiscent
of the envelope in figure~\ref{fig:numsims2}.
\label{fig:humps}}
\end{figure}

We now wish to determine the effect of small $\delta$ upon the envelope
of the pattern, since our numerical simulations suggest that the envelope
will broaden from a sech profile as the envelope becomes more developed.
For small $\delta$, (\ref{eq:Adel}) is approximately
\beq
0=(1-M)A+A_{XX}-(1-\mu'-\delta M/\sigma)A|A|^2-\mu'\delta A|A|^4/\sigma.
\label{eq:quint}
\eeq
and so to determine the effect of $\delta$ upon the envelope,
it is instructive to examine the model equation
\beq
f''=f-f^3+\delta f^5.
\label{eq:fmod}
\eeq
This model contains the essential features of the problem (\ref{eq:Mint}) and
(\ref{eq:quint}) for $A$, and has the exact solution
\beq
f=b^{1/2}\left(\tsfrac14(b-2)+\cosh^2x\right)^{-1/2},
\label{eq:fsol}
\eeq
where $ b = 2 / (1 - 16 \delta /3 )^{1/2}$.
When $\delta=0$, $f=\sqrt2\sech \,x$, but as $\delta$ increases the
profile broadens and flattens (see figure~\ref{fig:humps}),
giving envelopes reminiscent of those in figure~\ref{fig:numsims2}.

\section{Examples}
\label{sec:examples}

Since conservation laws arise very commonly in nature, 
it turns out that a wide variety of pattern formation problems are
governed by the system (\ref{eq:AT})--(\ref{eq:BT}) rather than the
Ginzburg--Landau equation alone. 

In thermal convection in a rotating layer with stress-free boundaries,
the conserved quantity is horizontal momentum and the corresponding
long-wave mode is a large-scale flow along the axes of the convection
rolls. In the analysis of Cox and Matthews (2000), 
(\ref{eq:AT})--(\ref{eq:BT}) appear as a special case, and it can be
shown that rotating convection is unstable for moderate values of the
Prandtl number. 

In Marangoni convection (driven by surface tension),  the fluid has a
free surface and the equation for the layer depth is constrained by
conservation of mass. Equations similar to (\ref{eq:AT})--(\ref{eq:BT}),
but with one extra term, were derived by 
Golovin, Nepomnashchy and Pismen (1994); in this case, however, 
the parameter $\mu$ is negative and so no instability arises.

In \S\S~\ref{sec:pensmall} and~\ref{sec:thermsmall} below, we derive an
equation similar to (\ref{eq:toypde2}) for two problems in
convection; in \S~\ref{sec:magsmall}, we directly derive amplitude
equations equivalent to (\ref{eq:AT})--(\ref{eq:BT}) for
magnetoconvection between `ideal' horizontal boundaries.

\subsection{Application to convection with a nonlinear equation of state}
\label{sec:pensmall}

A physical problem that gives rise to amplitude equations of the form
considered here is convection between boundaries at which the heat flux is
fixed (Roberts 1985, Matthews 1988a, 1988b). The fluid has a
nonlinear equation of state,
\beq
\rho_*=\rho_0\left[1-\alpha(T_*-T_0)-\beta(T_*-T_0)^2\right],
\label{eq:nlstate}
\eeq
relating the fluid density $\rho_*$ to the temperature $T_*$.
The coefficients $\alpha$, $\beta$ and $T_0$ are constants.
This equation of state is appropriate for water around $4^\circ$C, for example.
The crucial feature of this problem is that the total heat in the layer
is conserved; the fixed-flux boundary
conditions allow a large-scale, slowly-evolving depth-averaged
temperature mode, $\Theta$.

We suppose that the temperature profile is nonlinear,
of the form (Matthews 1988a, 1988b):
\beq
T_*=T_0+{\mathcal A}z_*^3-{\mathcal B}z_*,
\label{eq:Tprof}
\eeq
where $z_*$ is the vertical coordinate. The problem is expressed in
dimensionless variables by adopting the scales
$d=({\mathcal B}/{\mathcal A})^{1/2}$,
${\mathcal B}d$ and $d^2/\kappa$ for length, temperature and time. We then find the
equations governing two-dimensional motions to be
\begin{eqnarray}
\prandtl^{-1}
\left(\nabla^2\psi_t+J(\psi,\nabla^2\psi)\right)
&=&R\theta_x+\nabla^4\psi+R\Delta(z^3-z+\theta)\theta_x,
\label{eq:psi}\\
\theta_t+J(\psi,\theta_z)+(3z^2-1)\psi_x&=&\nabla^2\theta,
\label{eq:th}
\end{eqnarray}
where the $x$- and $z$-components of velocity may be expressed in terms
of the streamfunction $\psi$ as $(u,w)=(-\psi_z,\psi_x)$, and the Jacobian is
defined by $J(f,g)=f_xg_z-f_zg_x$.
The perturbation to the temperature profile (\ref{eq:Tprof}) is denoted
by $\theta$.
The Rayleigh number is
\[
R=\frac{g\alpha {\mathcal B}d^4}{\nu\kappa},
\]
where $\nu$ and $\kappa$ are, respectively, the coefficient of kinematic
viscosity and the coefficient of thermal diffusivity, $g$ is the
gravitational acceleration;
the Prandtl number is $\prandtl = \nu/\kappa$,
and $\Delta=2{\mathcal B}d\beta/\alpha$. In deriving the PDE for $\Theta$, we
assume that the equation of state is only mildly nonlinear,
so that $|\Delta|\ll1$.

We break the up--down symmetry of the problem by imposing rigid boundary
conditions at $z=-H$ but stress-free conditions at $z=H$. Thus
\beq
\psi=\psi_z=0 \quad\mbox{at }z=-H
\label{eq:bc-H}
\eeq
and
\beq
\psi=\psi_{zz}=0 \quad\mbox{at }z=H.
\label{eq:bcH}
\eeq
The important mathematical consequence of these boundary
conditions is that the equation for $\Theta$ then contains a quadratic
nonlinear term of the form $(\Theta_X^2)_{XX}$, where $X$ is a scaled
$x$-coordinate.
This term tends to lead to a supercritical bifurcation to rolls,
unlike the quadratic term $(\Theta^2)_{XX}$, which arises from the
nonlinear equation of state, and which tends to lead to a
subcritical bifurcation.

\begin{figure}
\epsfxsize0.5\hsize\epsffile{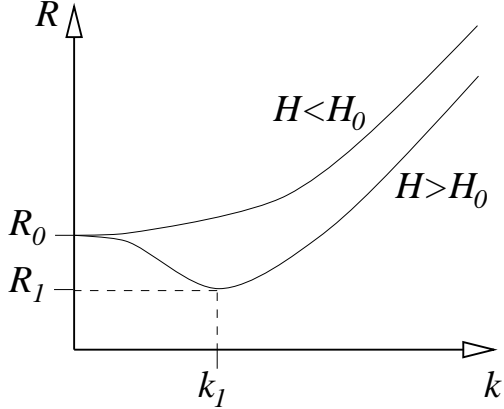}
\caption{Marginal stability curves for convection with a nonlinear
equation of state and nonlinear basic-state temperature profile.
When the depth $H$ of the fluid layer is less than some threshold $H_0$,
the critical wavenumber is zero and the critical Rayleigh number
is $R_0$, given by (\ref{eq:grow}); when $H>H_0$, the critical wavenumber
$k_1$ is positive, and the critical Rayleigh number $R_1$ is less than $R_0$.
\label{fig:HH0}}
\end{figure}

We suppose that the heat flux through the planes $z=\pm H$ is fixed,
so that the perturbation heat flux $\pa\theta/\pa z=0$ at each boundary.
The key consequence of this choice of boundary condition is
that, for sufficiently small half-depth
$H$, the critical mode of the linear stability problem has zero
wavenumber (see figure~\ref{fig:HH0}).

We begin by considering the simpler case of a linear equation of state,
with $\Delta=0$. First we examine equations (\ref{eq:psi})--(\ref{eq:th}),
linearised about the trivial state $\psi=\theta=0$.
We find, making asymptotic expansions of the forms given by
Roberts (1985) and Matthews (1988a), that
modes with wavenumber $k$ have linear growth rate $\lambda$, where
\begin{equation}
\lambda\sim (R-R_0)R_0^{-1}k^2-a_4k^4,\qquad R_0=420H^{-4}(11H^2-21)^{-1},
\label{eq:grow}
\end{equation}
for small $|R-R_0|$ and $k$. The coefficient $a_4$ is
\[
a_4=\frac{8H^2(6831H^4-25738H^2+23751)}{1287(11H^2-21)^2}.
\]
As a consistency check on these results, we note that as $H\to0$,
$R_0\sim20/H^4$ and $a_4\sim232H^2/693$. This limit corresponds to the
problem with a linear equation of state and linear basic-state temperature
profile, such as considered by Chapman and Proctor (1980).
After appropriately rescaling of the depth of our fluid layer, we find that
their results agree with ours in the limit as $H\to0$.
When $a_4>0$, the largest wavenumbers in (\ref{eq:grow}) are damped, and
$k=0$ corresponds to a minimum of the marginal stability curve. This
condition is satisfied for $0<H<H_0$, where $H_0\approx1.2709$ is the smaller
positive root of $a_4=0$ (see figure~\ref{fig:HH0}).

The amplitude equation of interest applies when $H$ is close to $H_0$.
We thus follow Roberts (1985) and Matthews (1988a) and let
\beq
R=R_0+\epsilon^2 R_2+\epsilon^4 R_4,\qquad
\Delta=\epsilon^2 D,\qquad
H=H_0+\epsilon^2 H_2,
\label{eq:pensca}
\eeq
expanding the streamfunction and temperature perturbation as
\begin{eqnarray}
\theta&=&\epsilon^2\left[\theta_2(X,T)+\epsilon^2\theta_4(X,z,T)+\cdots
                   \right]\label{eq:thexp}\\
\psi&=&\epsilon^3\left[\psi_3(X,z,T)+\epsilon^2\psi_5(X,z,T)+\cdots
                 \right],
\end{eqnarray}
where $X=\epsilon x$ and $T=\epsilon^6 t$. These scalings have
been adopted to ensure a balance of terms in the equation derived
below. Of course, it is reasonable to question the physical
relevance of such a large power of $\epsilon$ in the scaling of
time, and we offer two partial justifications. The first is that,
in fact, solutions to model PDEs of the form to be derived below
are known to provide a good approximation to the full convection
problem even when $\epsilon$ is not particularly small (and hence
$\epsilon^6$ represents a physically achievable scaling). The
second justification is that in deriving amplitude equations of
the form (\ref{eq:AT})--(\ref{eq:BT}) it is not {\em necessary} to
pass through an intermediate stage involving the time scale
$\epsilon^6 t$, this is just a convenient way to construct {\em in
an asymptotic fashion} a long-wavelength model PDE.

At $O(\epsilon^2)$ in (\ref{eq:psi})--(\ref{eq:th}), we find that
$\partial^2\theta_2/\partial z^2=0$, subject to the boundary conditions
$\partial\theta_2/\partial z=0$ at $z=\pm H$. The solution for $\theta_2$
is therefore independent of depth, as indicated in (\ref{eq:thexp}).
After some algebra, we find from a solvability condition at $O(\epsilon^8)$
that the amplitude equation for $\Theta=\theta_2(X,T)$ is
\beq\fl
\Theta_T=\left\{\left(-\frac{R_4}{R_0}+a_2D^2\right)\Theta+
(a_4'H_2+a_4''D)\Theta_{XX}+a_6\Theta_{XXXX}
+a_q\Theta_X^2-\half D \Theta^2\right\}_{XX}
\label{eq:412}
\eeq
where the numerical values of the coefficients in (\ref{eq:412})
are given approximately by
\begin{eqnarray}
a_2&=&7.588\times 10^{-5},\\
a_4'&=&8.966,\\
a_4''&=&0.1208,\\
a_6&=&0.3067,\\
a_q&=&5.571-1.086/\prandtl,\\
R_0&=&49.80.
\end{eqnarray}
To ensure that all terms in the equation for $\Theta_T$ arise at
the same asymptotic order, $R_2=a_7D$, where
$a_7\approx0.4338$.

After rescaling, we can write (\ref{eq:412}) as 
\beq
\frac{\partial w}{\partial t}  =  -\frac{\partial^2}{\partial x^2}
\left[ r  w - \left(1+ \frac{\partial^2}{\partial x^2} \right)^2 w
           -  s w^2  - \left(\prt w x \right)^2  \right],\label{eq:toypde}
\eeq
which is very similar in form to (\ref{eq:toypde2}) considered in 
\S~\ref{sec:toypde}.
Following the asymptotic procedure outlined in
\S~\ref{sec:toypde}, we find that with $w(x,t)$ as in
(\ref{eq:wexp}), the amplitudes of the pattern and large-scale
modes satisfy
\bea
A_T & = & r_2 A + 4 A_{XX} - 2(1-s)(s+2)A|A|^2/9 - 2 s AB  ,
\label{eq:AToy}\\ B_T & = & B_{XX} + 2(s+1) (|A|^2)_{XX}.
\label{eq:BToy}
\eea

If $-2<s<1$, the coefficient of the term $A|A|^2$ in
(\ref{eq:AToy}) is negative, and the bifurcation to pattern (in
the absence of $B$) is supercritical. Equations (\ref{eq:AToy})
and (\ref{eq:BToy}) are then equivalent to
(\ref{eq:AT})--(\ref{eq:BT}), after an appropriate rescaling.

From this analysis  we find that the bifurcation to rolls is
supercritical if $-2<s<1$, and that all rolls are unstable if
\beq\fl
-2<s<-\frac12-\frac{3\sqrt{57}}{38}\approx-1.096
\quad\mbox{or}\quad
0.096\approx-\frac12+\frac{3\sqrt{57}}{38}<s<1,
\label{eq:alluns}
\eeq
where
\beq
s=\frac{a_6^{1/2}R_0^{1/2}\Delta}{2a_q(R_0+a_7\Delta+a_2R_0\Delta^2-R)^{1/2}}.
\eeq

\subsection{Application to thermosolutal convection}
\label{sec:thermsmall}

As a second physical example, we examine thermosolutal convection between
boundaries at which the heat flux is fixed.
We suppose that the fluid is slightly non-Boussinesq, so that the
thermal diffusivity is a weak function of the local
temperature:
\beq
\kappa_*=\kappa_0(1+\hat\kappa(T_*-T_0)/T_0).
\eeq
The layer is heated from below and in the basic state the saltier water
lies over the fresher water, so both mechanisms are destabilising.
Both the thermal and the solutal gradients are vertical and uniform in
the absence of convection. The
thermal gradient is characterised by a (thermal) Rayleigh number $R>0$
and the salinity gradient by a solutal Rayleigh number $S<0$. When
$0>S>S_c$,
the bifurcation to convection takes place at zero wavenumber;
when $S_c>S$, the critical wavenumber is finite. We analyse the case
\[
S=S_c+\epsilon^2 S_2.
\]

The natural scale for lengths is now the depth of the layer,
and that for temperature perturbations is the temperature difference across
the layer in the basic state (similarly for the salt concentration).
With these scales, the dimensionless governing equations may be written
in the form
\bea
\prandtl^{-1}\left(\nabla^2\psi_t+J(\psi,\nabla^2\psi)\right)&=&
R\theta_x-S\Sigma_x+\nabla^4\psi\label{eq:ts1}\\
\theta_t+J(\psi,\theta)-\psi_x&=&\nabla^2\theta+
\hat\kappa\vec\nabla\cdot\left[(-z+\theta)\vec\nabla\theta\right]\\
\Sigma_t+J(\psi,\Sigma)-\psi_x&=&\tau\nabla^2\Sigma,\label{eq:ts3}
\eea
where $\Sigma$ is the perturbation salinity field, $\tau=\kappa_s/\kappa$
and $\kappa_s$ is the salt diffusivity.
The thermal Rayleigh number is now
\beq
R=\frac{g\alpha({\rm d}T_*/{\rm d}z_*)d^4}{\nu\kappa_0},
\eeq
where $\alpha$ is the coefficient of cubical expansion and
${\rm d}T_*/{\rm d}z_*$ is the (dimensional) temperature gradient in the
absence of convection; the solutal Rayleigh number $S$ is defined similarly.

As in the previous example, we use free--rigid horizontal boundary
conditions, with (\ref{eq:bc-H}) and (\ref{eq:bcH}) applied for $H=1/2$,
since the fluid layer now has unit dimensionless depth.
With these boundary conditions, $S_c=-8352\tau/41\approx-203.707\tau$.

The amplitude equation of interest applies when
\[
R=R_0+\epsilon^4 R_4,\qquad
S=S_c+\epsilon^2S_2,\qquad
\hat\kappa=\epsilon^2\kappa_2;
\]
we expand the independent variables as
\begin{eqnarray}
\theta&=&\epsilon^2\left[\theta_2(X,T)+\epsilon^2\theta_4(X,z,T)+\cdots
                   \right]\\
\Sigma&=&\epsilon^4\left[\Sigma_4(X,z,T)+\epsilon^2\Sigma_6(X,z,T)
                                                            +\cdots
                   \right]\\
\psi&=&\epsilon^3\left[\psi_3(X,z,T)+\epsilon^2\psi_5(X,z,T)+\cdots
                 \right],
\end{eqnarray}
where $X=\epsilon x$ and $T=\epsilon^6 t$.
At $O(\epsilon^8)$ in an asymptotic expansion of (\ref{eq:ts1})--(\ref{eq:ts3}),
we find that
the amplitude equation for $\Theta=\theta_2(X,T)$ is
\beq
\Theta_T=\left\{-\frac{R_4}{R_0}\Theta+c_4S_2\Theta_{XX}+
c_6\Theta_{XXXX}+c_q\Theta_X^2 +\half \kappa_2\Theta^2\right\}_{XX}.
\eeq
where
\begin{eqnarray}
R_0&=&320,\\
c_4&=&-\frac{41}{99792\tau}\approx-0.0004109/\tau,\\
c_6&=&\frac{14506722953}{1105865601750}\approx0.01312,\\
c_q&=&-\frac{5+7\prandtl}{84\prandtl}\approx-0.08333-0.05952/\prandtl.
\end{eqnarray}
Note that this has exactly the same form as (\ref{eq:toypde}).
From \S~\ref{sec:pensmall}, the condition for a supercritical bifurcation is
$-2<s<1$, and that for all rolls to be unstable is (\ref{eq:alluns}), where
\beq
s=\frac{\hat\kappa c_6^{1/2}R_0^{1/2}}{2c_q(R_0-R)^{1/2}}.
\eeq


\subsection{Application to magnetoconvection}
\label{sec:magsmall}

Convection in a vertical magnetic field has been a subject of a great deal of
research (see, for example, Proctor and Weiss 1982; Matthews 1999).
The problem is motivated by the convection zone of the Sun,
where the magnetic field is moved around by the fluid
and regions of strong magnetic field inhibit the fluid motion.
In the nonlinear regime magnetoconvection exhibits a very wide range
of complicated phenomena.

In magnetoconvection, the conservation law that leads to the
large-scale mode is that the total flux of
magnetic field through the layer is conserved.
The governing equations for two-dimensional magnetoconvection
are (Proctor and Weiss 1982), in the notation of the preceding section,
\bea
\fl\prandtl^{-1}\left(\nabla^2\psi_t+J(\psi,\nabla^2\psi)\right)  &=&
       R \, \theta_x + \nabla^4\psi + \zeta Q \left( J(\phi,\nabla^2\phi)
       +\nabla^2\phi_z\right) ,   \label{eq:magnse} \\
\theta_t+J(\psi,\theta)  &=&  \psi_x + \nabla^2\theta  , \label{eq:magheat}\\
\phi_t+J(\psi,\phi)  &=&  \psi_z + \zeta \nabla^2\phi  \label{eq:magmag}.
\eea
The magnetic field is defined through a perturbation flux function $\phi$
by $(B_x, B_z) = (-\phi_z, 1 + \phi_x)$.
The parameter $Q$ is the Chandrasekhar number, a dimensionless measure
of the magnetic field strength, while $\zeta$ is the ratio of magnetic
diffusivity to thermal diffusivity.

The most commonly used boundary conditions are that the upper and
lower boundaries are stress free, fixed temperature and with no
horizontal magnetic field, so
\beq
\psi = \psi_{zz} = \theta = \phi_z = 0 \quad \mbox{at} \quad z=0,1 .
\eeq
These boundary conditions are convenient in the sense that the
linear eigenfunctions are trigonometric.

The large-scale mode in the problem arises through a rearrangement of
the vertical magnetic field. There is a linear eigenfunction in which
$\phi = \phi(x,t)$ which obeys the diffusion equation
$\phi_t = \zeta \phi_{xx}$, so the growth rate $\lambda$ of a mode
with wavenumber $k$ is simply $\lambda = - \zeta k^2$.
Note that this mode is independent of the velocity and temperature
perturbation,  since if $\phi = \phi(x,t)$ there is no forcing term
in (\ref{eq:magnse}). Although this mode is usually ignored in the
analysis of magnetoconvection, it plays a crucial role when the
horizontal size of the domain is large.
Note that although there are nearly marginal modes at small $k$ and at
order-one values of $k$, the graph of growth rate versus wavenumber is
not as shown in figure~\ref{fig:lambda}, because the large-scale mode
and the order-one mode correspond to two distinct branches of eigenfunctions.

The weakly nonlinear analysis proceeds by including the large-scale
mode. The general analysis of \S 2 suggests that the mean mode should
be included at order $\ep^2$. In fact the mean vertical magnetic field
is of order $\ep^2$ and so the corresponding flux function $\phi$ is of
order $\ep$.
The appropriate asymptotic expansions near the stationary bifurcation to
convection are therefore
\bea
\psi   & = & \ep A(X,T) \exp \i kx \sin \pi z + c.c. + O(\ep^2), \\
\theta & = & \ep A(X,T)\,\theta_1 \i \exp \i kx \sin \pi z + c.c. + O(\ep^2),\\
\phi   & = & \ep A(X,T)\,\phi_1  \exp \i kx \cos \pi z + c.c.
           + \ep \, \phi_1 \, \Phi(X,T) + O(\ep^2),
\eea
where $\theta_1$ and $\phi_1$ are constants, readily determined from the
linear stability problem, $X = \ep x$, $T = \ep^2
t$ and $R = R_c + \ep^2 R_2$.
At order $\ep$ we obtain $\theta_1 = k/(\pi^2+k^2)$,
$\phi_1 = \pi/\zeta(\pi^2+k^2)$ and the well known result
(Chandrasekhar 1961)
\beq
R_c = \frac {(\pi^2 + k^2)^3 + Q \pi^2 (\pi^2 + k^2)}{k^2} \label{eq:magrc}.
\eeq
From (\ref{eq:magrc}) it follows that as $R$ is increased, convection
first occurs with a critical wavenumber $k_c$ which is related to $Q$
by
\beq
Q \pi^4 = (2 k_c^2 - \pi^2) (\pi^2 + k_c^2)^2  \label{eq:magkc}.
\eeq
Henceforth it is assumed that $k=k_c$; note that the envelope function
$A(X,T)$ permits convection with a wavenumber $k=k_c + O(\ep)$.

At order $\ep^2$, a term in $\theta$ proportional to $\sin 2 \pi z$
and a $\phi$ term  proportional to $\exp 2 \i k x$ are obtained.
At third order, evolution equations for $A$ and $\Phi$ are found
by applying solvability conditions.
Full details of the calculation will be given elsewhere.
The resulting amplitude equations are
\bea
A_T & = & a_1 A + a_2 A_{XX} - a_3 A|A|^2 - a_4 A \,\Phi_X  ,\label{eq:magAT}\\
\Phi_T & = & \zeta \Phi_{XX}  + \pi \left(|A|^2\right)_{X} , \label{eq:magBT}
\eea
where $a_1 \ldots a_4$ are known constants, given by
cumbersome functions of the parameters; $a_2$ and $a_4$ are always
positive.
After setting $\Phi_X = B$, (\ref{eq:magAT})--(\ref{eq:magBT})
are equivalent to (\ref{eq:AT})--(\ref{eq:BT}).
The equations (\ref{eq:magAT})--(\ref{eq:magBT}) have not been derived
previously, but the underlying processes of the coupling between
the convection rolls $A$ and the large-scale mode $\Phi$ are well
understood (see, for example, Proctor and Weiss 1982).
Since $a_4 > 0$, the coupling term in (\ref{eq:magAT}) simply reflects
the fact that the magnetic field suppresses convection.
The coupling term in (\ref{eq:magBT}) describes the process of flux
expulsion: the magnetic field strength decreases in regions where the
convection is strong.

\begin{figure}

\epsfxsize0.8\hsize\epsffile{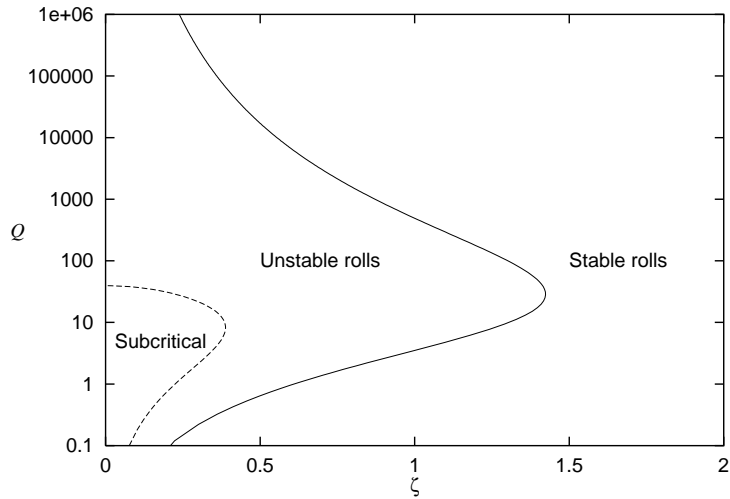}
\caption{Region of stability of convection rolls in magnetoconvection.
To the right of the solid line, rolls are stable.
To the left of the dashed line the bifurcation is subcritical.
Between the two lines, all rolls are unstable. }
\label{fig:magstab}
\end{figure}

The bifurcation at the onset of convection is supercritical (i.e.\ $a_3 > 0$)
provided that
\beq
k^4 \zeta^2 (\pi^2+k^2) > \pi^2 (2k^2 - \pi^2)(\pi^2 - k^2),\label{eq:magsup}
\eeq
where $Q$ has been eliminated in favour of $k$ using (\ref{eq:magkc}).
The result (\ref{eq:stabcond}) can now be applied to determine whether
a regular pattern of rolls is unstable. It is found that all rolls are
unstable if
\beq
k^4 \zeta^2 (\pi^2+k^2) < \pi^2 (2k^2 - \pi^2)(k^2 + 3\pi^2),\label{eq:magstab}
\eeq
where $Q$ has again been eliminated using (\ref{eq:magkc}).
Note that neither of the conditions (\ref{eq:magsup})--(\ref{eq:magstab})
involves the Prandtl number $\prandtl$.
Using $k$ as a parameter, the region of stability of rolls can be
plotted in the $Q$, $\zeta$ plane (figure~\ref{fig:magstab}).
In the small region where the rolls are subcritical the weakly
nonlinear analysis gives no useful information about the pattern.
However, there is a large region of parameter space in which the
bifurcation is supercritical but all rolls are unstable.
In this region we expect to find that the stable pattern at onset
consists of rolls that are modulated on a large length-scale.
Solutions of this type, but fully nonlinear, have recently
been obtained through numerical simulations of magnetoconvection
(Blanchflower 1999).

Note that our analysis is confined to the onset of convection at
a stationary bifurcation.
For certain parameter values (in general, when $Q$ is large and
$\zeta$ is small) the onset of magnetoconvection is
oscillatory; since this depends on the value of $\prandtl$, this
condition is not shown in figure~\ref{fig:magstab}.

\section{Discussion}
\label{sec:disc}

We have derived amplitude equations (\ref{eq:AT})--(\ref{eq:BT})
for the primary bifurcation to a stationary 
pattern in a one-dimensional system with a conservation law for a
quantity that does not change sign under reflection.
In this case the conservation law leads to a neutral large-scale mode
which must be taken into account for a correct description of the 
behaviour near onset. 
The usual Ginzburg--Landau equation (Melbourne 1999) is coupled to an
equation describing the large-scale evolution of the conserved quantity.
We have shown how these equations may be obtained from symmetry
considerations alone, and have also derived them directly from the
governing equation(s) of the system.

In considering the small-amplitude solutions of the amplitude equations,
we have shown how periodic patterns may become unstable to a new
type of instability. This new instability is quite different from the 
usual Eckhaus instability in three ways. It is an 
amplitude-driven instability (whereas the Eckhaus instability is
phase-driven) and it is in general supercritical.  
Most importantly, for certain parameter values {\it all} patterns are unstable.
Our instability is also different from that which occurs in the case
where the conserved quantity is velocity-like, and
the reflection symmetry acts as $-1$ (such as in systems with
Galilean invariance), where periodic patterns are always unstable and
chaotic dynamics is observed at onset (Tribelsky and Velarde 1996,
Tribelsky and Tsuboi 1996, Matthews and Cox 2000).

Beyond the instability, there are stable solutions in the form of a
strongly amplitude-modulated periodic pattern, the envelope of which
we have calculated in terms of Jacobi elliptic functions. The envelope
may commonly be approximated by a sech profile.
Numerical simulations reveal that these strongly modulated patterns may be
stable in the full governing PDE, not just within the framework of the
amplitude equations. Numerical simulations of the governing PDE
also demonstrate that the amplitude equations represent a good approximation
provided the amplitude is sufficiently small,
but that higher-order terms play an important role 
as the amplitude increases; these higher-order terms
can lead to significant modifications in the bifurcation diagram,
and a broadening of the envelope of the pattern.

Our work demonstrates that localised solutions are possible when a
Ginzburg--Landau equation is coupled to an equation for a mean field,
even when the coefficients of the equations are real and when the bifurcation
is supercritical. Previous work has required complex coefficients
(Riecke 1992a, 1992b, Sakaguchi 1993), a subcritical bifurcation (Herrero and
Riecke 1995, Sakaguchi and Brand 1996) or both.
Our amplitude equations apply when the bifurcation from the
basic state is stationary;  the case of oscillatory bifurcation to
travelling-wave solutions has been analysed in a series of papers
by Riecke and co-workers (Riecke 1992a, 1992b, 1996, Herrero and Riecke 1995)
and by others (Sakaguchi 1993, Barthelet and Charru 1998),
who find a variety of localised and front-type solutions.
They find that the coupling to a mean field (our $B$) can have
a significant effect on solutions and their stability, and can stabilise
localised travelling-wave trains, and holes.

We have applied our amplitude equations to three convection problems,
leading to novel instabilities in these systems.
Further details of these applications will be given elsewhere. 
In the first two, the heat flux is fixed on the boundaries, and the
total heat in the layer is conserved, which leads to a large-scale
mode in the form of a depth-averaged temperature perturbation.
However, in order for our amplitude equations
(\ref{eq:AT})--(\ref{eq:BT}) to apply, this large-scale mode must
influence the convection rolls; this is achieved  by using either a
nonlinear equation of state or a temperature-dependent diffusivity.
In the third example, magnetoconvection, the conserved quantity is the
vertical magnetic field; in this case 
the large-scale mode and the pattern mode correspond to two distinct
branches of eigenfunctions of the linear problem.
Our modulated solutions may be related to the isolated
`convectons' found recently in magnetoconvection (Blanchflower 1999).
Another application of our work is in rotating convection, 
where the conserved quantity is momentum (Cox and Matthews 2000).

Our amplitude equations (\ref{eq:AT})--(\ref{eq:BT}) are identical to those
derived by Coullet and Iooss (1990), who considered the {\em secondary}
stability of a one-dimensional cellular pattern. In that case, $A$ corresponds
to the amplitude of a perturbation to the basic pattern, and $B$
to a gradient in the phase of the pattern; the amplitude equations then
describe evolution near a stationary bifurcation.
Coullet and Iooss (1990) found that when a pattern of wavelength $L$ becomes
unstable to a perturbation of wavelength $2L$, the amplitude equations
(\ref{eq:AT})--(\ref{eq:BT}) apply, with $A$ real; when the spatial
period of the perturbation and the basic pattern are irrationally related,
$A$ may be complex.
Our results suggest that if a pattern of wavelength $L$ undergoes a
supercritical subharmonic bifurcation then there may be stable states
where the original pattern is largely undisturbed in some regions but where
the perturbation is strong in others.
Furthermore, provided the secondary bifurcation is supercritical,
it should be possible to find stable states consisting of two domain types:
in one domain the pattern is essentially the basic cellular state,
in the other the pattern is strongly perturbed.

It would be of interest to investigate how this theory may be extended
to problems in two or three space dimensions; in particular we would
like to know whether solutions may be obtained that are localised in space.
Such localised solutions have been found experimentally by VanHook 
{\it et al.}\  (1995, 1997).
Localised solutions after an oscillatory bifurcation have recently been
examined in two dimensions by Riecke and Granzow (1998), who found that
such solutions may be obtained in a pair of Ginzburg--Landau equations coupled
to a mean-field equation. Of particular interest is their result that
stable localised states can be found, as here, when the primary bifurcation
is supercritical and the coefficients of the amplitude equations are real.

\ack{We have benefited from useful discussions with Richard Tew on the subject
of Jacobi elliptic functions.}


\References

\item{}
Abramowitz  M and  Stegun I A
1965
{\it Handbook of Mathematical Functions}
(New York: Dover).

\item{}
Barthelet P and Charru F 
1998
Benjamin--Feir and Eckhaus instabilities with Galilean invariance:
the case of interfacial waves in viscous shear flows
{\it Eur. J. Mech.  B} {\bf 17} 1--18.

\item{}
Blanchflower S
1999
Magnetohydrodynamic convectons
{\it Phys.\ Lett.\ A} {\bf 261} 74--81.

\item{}
Chandrasekhar S
1961
{\it Hydrodynamic and hydromagnetic stability}
(New York: Dover).

\item{}
Chapman C J and Proctor M R E
1980
Nonlinear Rayleigh--B\'enard convection between poorly conducting boundaries
{\it J. Fluid Mech.}
{\bf 101}
759--782.

\item{}
Charru F and Barthelet P 
1999
Secondary instabilities of interfacial waves due to coupling with a
long wave mode in a two-layer Couette flow
{\it Physica}
{\bf D 125}
311--324.

\item{}
Cox S M
1998
Long-wavelength rotating convection between poorly conducting boundaries
{\it SIAM J. Appl. Math.}
{\bf 58}
1338--1364.

\item{}
Cox S M and Matthews P C
2000
Instability of rotating convection
{\it J. Fluid Mech.}
{\bf 403}
153--172.

\item{}
Cross M C and Hohenberg P C
1993
Pattern formation outside of equilibrium
{\it  Rev.\ Mod.\ Phys.}
{\bf 65}
851--1112.

\item{}
Coullet P and  Fauve S
1985
Propagative phase dynamics for systems with Galilean invariance
\PRL
{\bf 55}
2857--2859

\item{}
Coullet P and  Iooss G
1990
Instabilities of one-dimensional cellular patterns
\PRL
{\bf 64}
866--869.

\item{}
Eckhaus W
1965
{\it Studies in Non-linear Stability Theory}
(New York: Springer).

\item{}
Fauve S
1987
Large scale instabilities of cellular flows.
In Instabilities and nonequilibrium structures,
ed. E. Tirapegui and D. Villarroel, 
63--88.

\item{}
Golovin A A, Nepomnashchy A A and Pismen L M
1994
Interaction between short-scale Marangoni convection and long-scale
deformational instability
{\it Physics of Fluids}
{\bf 6} 34--48.

\item{}
Herrero H and Riecke H
1995
Bound pairs of fronts in a real Ginzburg--Landau equation coupled
to a mean field
{\it Physica}
{\bf D 85}
79--92.

\item{}
Hidaka Y, Huh J-H, Hayashi K-i, Kai S and Tribelsky M I 1997
Soft-mode turbulence in electrohydrodynamic convection of a
homeotropically aligned nematic layer {\it Phys. Rev E} {\bf 56}
R6256--R6259.

\item{}
Matthews P C 1988a Convection and mixing in ice-covered lakes (PhD
thesis, University of Cambridge).

\item{}
Matthews P C
1988b
A model for the onset of penetrative convection
{\it J. Fluid Mech.}
{\bf 188}
571--583.

\item{}
Matthews P C
1998
Hexagonal patterns in finite domains
{\it Physica}
{\bf D 116}
81--94.

\item{}
Matthews P C
1999
Asymptotic solutions for nonlinear magnetoconvection
{\it J. Fluid Mech.}
{\bf 387}
397--409.

\item{}
Matthews P C and Cox S M
1999
Pattern formation with Galilean invariance
{\it submitted}.

\item{}
Melbourne I
1999
Steady-state bifurcation with Euclidean symmetry
{\it Trans.\ Amer.\ Math.\ Soc.}
{\bf 351}
1575--1603.

\item{}
Proctor  M R E  and  Weiss N O
1982
Magnetoconvection
{\it Rep.~Prog.~Phys.}
{\bf 45}
1317--1379.

\item{}
Riecke H
1992a
Self-trapping of traveling-wave pulses in binary mixture convection
{\it Phys. Rev. Lett.}
{\bf 68}
301--304.

\item{}
Riecke H
1992b
Ginzburg--Landau equation coupled to a concentration field in
binary-mixture convection
{\it Physica}
{\bf D 61}
253--259.

\item{}
Riecke H
1996
Solitary waves under the influence of a long-wave mode
{\it Physica}
{\bf D 92}
69--94.

\item{}
Riecke H and Granzow G D
1998
Localization of waves without bistability: worms in nematics electroconvection
{\PRL}
{\bf 81}
333--336.

\item{}
Roberts A J
1985
An analysis of near-marginal, mildly penetrative convection with
heat flux prescribed on the boundaries
{\it J. Fluid Mech.}
{\bf 158}
71--93.

\item{}
Sakaguchi H and Brand  H R
1996
Stable localized solutions of arbitrary length for the quintic
Swift--Hohenberg equation
{\it Physica}
{\bf D 97}
274--285.

\item{}
Swift J B and Hohenberg P C
1977
Hydrodynamic fluctuations at the convective instability
{\it Phys. Rev. A} {\bf 15}
319--328.

\item{}
Tribelsky M I and Velarde M G
1996
Short-wavelength instability in systems with slow long-wavelength
dynamics. {\it Phys.  Rev.  E}
{\bf 54}  4973--4981.

\item{}
Tribelsky M I and Tsuboi K
1996
New scenario for transition to turbulence?
{\PRL}
{\bf 76} 1631--1634.

\item{}
Tuckerman L S and Barkley D
1990
Bifurcation analysis of the Eckhaus instability
{\it Physica} {\bf D 46} 57--86.

\item{}
VanHook S J, Schatz M F, McCormick W D, Swift J B and Swinney H L
1995 Long-wavelength instability in surface-tension-driven
B\'enard convection {\it Phys. Rev. Lett.} {\bf 75} 4397--4400.

\item{}
VanHook S J, Schatz M F, Swift J B, McCormick W D,  and Swinney H L 
1997 Long-wavelength surface-tension-driven B\'enard convection:
experiment and theory
{\it J. Fluid Mech.} 
{\bf 345} 45--78.

\item{}
Whittaker E T and Watson G N 1962 {\it A course of modern
analysis} (Cambridge University Press).

\endrefs

\end{document}